\newcommand{\be}{\begin{equation}}
\newcommand{\ee}{\end{equation}}

\documentclass[journal,10pt]{IEEEtran}

\usepackage{booktabs}
\usepackage{psfrag}
\ifCLASSINFOpdf
   \usepackage[pdftex]{graphicx}

\else
  
   \usepackage[dvips]{graphicx}

\fi
\usepackage[cmex10]{amsmath}
\usepackage{amsthm,epstopdf,nicefrac,textcomp}

 \usepackage{tikz} 
\usepackage{mathtools}

\usepackage[caption=false,font=footnotesize]{subfig}

\renewcommand\appendix{\par
  \setcounter{section}{0}
  \setcounter{subsection}{0}
  \setcounter{figure}{0}
  \setcounter{table}{0}
  \renewcommand\thesection{Appendix \Alph{section}}
  \renewcommand\thefigure{\Alph{section}\arabic{figure}}
  \renewcommand\thetable{\Alph{section}\arabic{table}}
}

\newcommand\copyrighttext{%
  \footnotesize \textcopyright 2014 IEEE.  Personal use of this material is permitted. Permission from IEEE must be obtained for all other uses, in any current or future media, including reprinting/republishing this material for advertising or promotional purposes, creating new collective works, for resale or redistribution to servers or lists, or reuse of any copyrighted component of this work in other works. 
 DOI: \href{https://doi.org/10.1109/TIM.2014.2341372}{10.1109/TIM.2014.2341372}
 }
\newcommand\copyrightnotice{%
\begin{tikzpicture}[remember picture,overlay]
\node[anchor=south,yshift=10pt] at (current page.south) {\fbox{\parbox{\dimexpr\textwidth-\fboxsep-\fboxrule\relax}{\copyrighttext}}};
\end{tikzpicture}%
}

\usepackage{color}

\twocolumn

\usepackage[bookmarks=false]{hyperref}

\begin{document}
\title{Information and Statistical Efficiency\\ When Quantizing Noisy DC Values}

\author{A.~Moschitta~\IEEEmembership{Member,~IEEE}\thanks{A. Moschitta is with the University of Perugia - Department DIEI, via G. Duranti, 93 - 06125 Perugia Italy,}
and~J.~Schoukens,~\IEEEmembership{Fellow Member,~IEEE}\thanks{J. Schoukens is with the Vrije Universiteit Brussel, Department ELEC, Pleinlaan 2, B1050 Brussels, Belgium.}
and P.~Carbone~\IEEEmembership{Senior Member,~IEEE}\thanks{P. Carbone is with the University of Perugia - Department DIEI, via G. Duranti, 93 - 06125 Perugia Italy.}}

\newtheorem{theorem}{Theorem}[section]
\newtheorem{lemma}[theorem]{Lemma}

\maketitle
\copyrightnotice
\begin{abstract}
\boldmath
%\today \newline
This paper considers estimation of a quantized constant in noise when using uniform and non--uniform quantizers. Estimators based on simple arithmetic averages, on sample statistical moments and on the maximum--likelihood procedure are considered. 
It provides expressions for the statistical efficiency of the arithmetic mean by comparing 
its variance to the proper Cram\'er--Rao lower bound. It is conjectured that the arithmetic
mean is optimal among all estimators with an exactly known bias. Conditions under which its
statistical performance are improved by the other estimation procedures when the exact bias is not known are found and analyzed. 
Using simulations and analysis of experimental data,
it is shown 
that both moment--based and maximum--likelihood--based 
estimators provide better results when the noise standard deviation is comparable to the quantization step and the {\em noise model} of quantization can not be applied.

\end{abstract}


%\psfrag{td}[lb][bl]{\small \hskip3cm $\theta/\Delta$}

\newcommand{\fg}[1]{{\frac{1}{\sqrt{2\pi}\sigma} e^{-\frac{{#1}^2}{2\sigma^2}}}} 

% \fi
\IEEEpeerreviewmaketitle

\section{Introduction}
When obtaining information about a constant measurand affected by 
noise with independent outcomes, 
the arithmetic mean is often taken as the simplest and 
most direct method to increase estimation accuracy and precision.
While this choice may be proven optimal under specific situations regarding the 
probability density function of the noise sequence, e.g. independent and equally distributed additive Gaussian noise, it may fail in many situations of 
practical relevance. 
This is the case for instance when data are quantized before the arithmetic
mean is performed. The quantization operation, being nonlinear, affects the noise
behavior so to invalidate the optimality of the mean value. Since quantization is applied
in the vast majority of engineered instruments and data acquisition boards, 
it is of interest to analyze its effects and to find alternative and more statistically efficient techniques to decode information about the measurand from the stream of quantized data.
While linearization of nonlinear systems and quantizers is a popular topic, 
when it comes to analog--to--digital (ADC) and digital--to--analog converters,
it requires a significant amount of information about the acquisition channel and
a major amount of computational resources, so that its usage is
thus confined to specific applications \cite{Schoukens}\cite{Handel0}.

The aim of this paper is twofold: showing how uniform 
quantization modifies the information available at the quantizer output 
for further processing and provide information for the user to take advantage of
this knowledge in order to adopt more efficient estimators. To achieve both goals a rigorous approach is taken in this paper to obtain expressions that are then simplified to obtain order of magnitudes and that can easily  be implemented in procedures to be used in practice. 

This topic was the subject of previous research. In \cite{Vardeman} a case study is analyzed to show the effect of quantization on the arithmetic mean of a sequence of rounded data.
The paper includes the analysis of the Likelihood Function (LF) of the observed sequence 
to find better estimates of the measurand and the corresponding confidence intervals.
In \cite{Gendai} the LF estimation is again adopted to estimate the transition levels in an 
analog--to--digital converter and the variance of the input--referred additive noise.
Maximum likelihood estimation based on quantized data is again considered in 
\cite{Gustafsson1} where the generation of {\em dither} noise is analyzed to smooth the LF.
The same authors, in \cite{Gustafsson2}, consider the identification of a quantized constant value in noise and provide general properties of the analyzed estimators.
In this paper we consider a similar measurement setup as in \cite{Gustafsson2}.
We first elaborate on the Cramer--Rao lower bound to find 
approximate upper and lower bounds to the mathematical expressions useful to show
the information loss caused by quantization. 
We conjecture that the canonical arithmetic mean
is optimal among all estimators having its estimation bias.
We than apply moment--based estimators and a likelihood estimator to quantized
data and determine the statistical efficiency with respect to simple average. 
Finally, we consider the quantizer affected by Integral Non Linearities (INL) and Differential Non Linearities (DNL) and show that the MLE still provides better results than the arithmetic mean under moderate amount of non uniformity in the position of transition levels.

\begin{figure}[h]
\begin{center}
\includegraphics[scale=0.65]{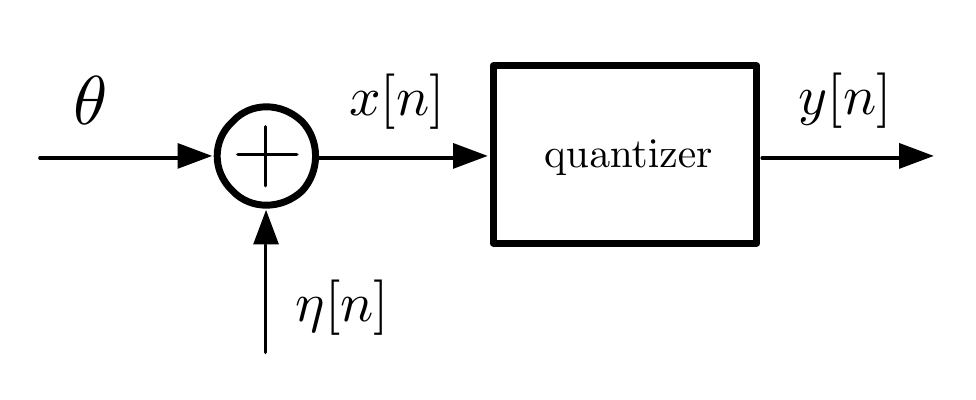}
\caption{The signal chain considered in this paper.\label{figchain}}
\end{center}
\end{figure}  

\section{Signals and systems}
The signal chain considered in this paper is shown in Fig.~\ref{figchain}. In this figure, 
$\theta$ represents the constant value to be estimated using 
repeated measurements of a quantized and noisy version of it.
In Fig.~\ref{figchain}, $\eta[\cdot]$ represents the zero--mean noise.
This is assumed as a sequence of 
Gaussian independent and equally distributed 
random variables with variance $\sigma^2$. 
A mid--tread {\em non--overloadable}
quantizer with input--output characteristic 
\be
	y[\cdot] = \Delta\left\lfloor \frac{x[\cdot]}{\Delta}+\frac{1}{2}\right\rfloor
\label{quant}
\ee
is considered, 
where $\Delta$ is the quantization step
and $\lfloor \cdot \rfloor$ represents the floor operator.
By defining $e[\cdot] \coloneqq y[\cdot]-x[\cdot]$ as the quantization error, we have: 
%When acquiring data using uniform quantization it is customary to use
%simple average to estimate mean values of input signals.
%Based on the criterion of the CRLB, in this subsection it is 
%assessed whether this approach is optimal or not.
%To this aim, consider that
\be
	y[n] = \theta+\eta[n]+e[n], \quad n=0, \ldots, N-1 
	\label{onebis}
\ee
It is common practice to adopt the {\em noise model} of quantization, 
that is to model the effect of quantization as an additive source of 
uniformly distributed random variables in the interval  $\left( -\frac{\Delta}{2}, \frac{\Delta}{2}\right)$ \cite{KollarBook}. While this approach might yield acceptable results under specific situations,
it fails if the dynamic of the input signal signal is not sufficiently large or in the presence of synchronous measurements 
where the `averaging' behavior is lost. Since, for small values 
of $\sigma^2$ this is not the case, the noise model of quantization is not expected to provide accurate results in our measurement conditions as will be shown in the next sections.

\subsection{The arithmetic mean estimator and its properties}

The arithmetic mean--based estimator $\hat{\theta}$ of $\theta$ is defined as
\be
	\hat \theta = \frac{1}{N}\sum_{n=0}^{N-1}y[n].
	\label{natest}
\ee
The bias and variance of (\ref{natest}) can be determined on the basis of the
first moments of the quantization error sequence. These, in turn, can be assessed
by using a frequency--domain approach to model the input--output 
characteristic of the quantizer. 
Since the statistical properties of the quantizer output and of the quantization error 
generally only depend on the ratio between $\sigma$ and $\Delta$, it is appropriate to     
define $\overline{\sigma} \coloneqq \nicefrac{\sigma}{\Delta}$. 
Then, in \cite{CarbonePetri}\cite{Carbone},  is shown that when
$\overline{\sigma} > 0.3$,
\be
	m_e(\theta) \coloneqq E\left\{ e[\cdot]\right\} \simeq -\frac{\Delta}{\pi} e^{-2\pi^2 \overline{\sigma}^2} \sin \left(2\pi \frac{\theta}{\Delta}\right), 
	%\qquad \frac{\sigma}{\Delta} \geq 0.3
	\label{me_gau}
\ee
\begin{align}
\begin{split}
\sigma_e^2(\theta) \coloneqq \mbox{var}(e[\cdot])&
%\frac{\serr(\nqe)}{\Delta^2}
		 	\simeq \frac{\Delta^2}{12}- \frac{\Delta^2}{\pi^2}e^{-2\pi^2 \overline{\sigma}^2}
			\left[
			   	\cos \left(2\pi \frac{\theta}{\Delta}\right) \right.\\  
		&  \left.	-e^{-2\pi^2\overline{\sigma}^2}\sin^2\left(2\pi \frac{\theta}{\Delta}\right)
			\right], %\qquad \frac{\sigma}{\Delta} \geq 0.3.
   \label{gau_var}
\end{split}
\end{align}
where $E\{\cdot\}$ is the expected value operator, and that for $\overline{\sigma} > 0.3$
\begin{align}
\begin{split}
\sigma_{y}^2(\theta) & \coloneqq \mbox{var}(y[\cdot])
%	\frac{\syns(\nqe)}{\Delta^2}
   		\simeq 
		\frac{\Delta^2}{12}+\sigma^2
   		\\
&   		-e^{-2\pi^2 
			\overline{\sigma}^2} \left[
			\left(
				4\sigma^2+\frac{\Delta^2}{\pi^2}
			\right)
			\cos\left( 2\pi \frac{\theta}{\Delta}\right) \right. \\
&   			\left. -\frac{\Delta^2}{\pi^2}
			e^{-2{\pi^2}\overline{\sigma}^2}
   			\sin^2\left(2\pi \frac{\theta}{\Delta}\right)
		\right]. %\qquad \frac{\sigma}{\Delta}  \geq 0.3,
\label{s2xq_gau}
\end{split}
\end{align}
Thus, from (\ref{natest}), we have:
\be
E({\hat \theta}) = \theta+m_e(\theta), 
\ee
and
\be
	\sigma_{{\hat \theta}}^2(\theta) \coloneqq  \mbox{var}({\hat \theta}) = \frac{\sigma_y^2(\theta)}{N} 
	\label{ssyq}
\ee
Expression (\ref{me_gau}) %--(\ref{s2xq_gau}) 
shows that the quantization error average
vanishes uniformly when $\overline{\sigma}$ increases. At the same time, 
the quantizer output variance achieves the value $\Delta^2/12+\sigma^2$ 
as expected when the noise model of quantization can be applied \cite{KollarBook}. 
Observe that the condition 
$\overline{\sigma} > 0.3$ in (\ref{me_gau})--(\ref{s2xq_gau}), 
occurs frequently when low--resolution ($<10$ bit) high--speed ADCs 
and medium/high--resolution low--speed ADCs are used.  Corresponding expressions
when $\overline{\sigma}< 0.3$ are available in \cite{KollarBook,CarbonePetri}.

\section{The Cramer--Rao Lower Bound of a Quantized Constant in Gaussian Noise}
%Thus, (\ref{natest}) is a biased estimator of 
%a quantized constant in noise and the CRLB must be modified to account for this bias.
The Cramer--Rao Lower Bound (CRLB) provides a limit on the minimum variance that can be attained by any estimator in the considered estimation problem.  
In the case of an estimator ${\hat \theta}$ of $\theta$ 
characterized by a bias $m_e(\theta)$, the CRLB becomes \cite{Kay}:
\be
CRLB_b(\theta) = \frac{\left(1+\frac{d m_e(\theta)}{d \theta}\right)^2}{I_N(\theta)}
	\label{crlbb}
\ee
where $I_N(\cdot)$ represents the Fisher information provided by $N$
samples of processed data \cite{Kay}. 
%Observe that when $\eta[\cdot] \equiv 0$, the quantizer output and error sequences become %deterministic functions of the input, $m_{e}(\theta)=\nicefrac{\Delta}{2}-\Delta\langle %\nicefrac{\theta}{\Delta} +\nicefrac{1}{2}\rangle$. Then, apart from a countable set of points,   
%$\nicefrac{dm_e(\theta)}{d \theta}=-1$ and $CRLB_b(\theta)=0$.

Given that the noise sequence has independent outcomes
$I_N(\theta)=N I_1(\theta)$ \cite{CarboneMoschitta}, where
% is provided by (\ref{fisher}). This bound is applicable to all
%estimators whose bias is $m_e(\theta)$. 
%Let us start with a single Gaussian observation.
%Assume
%\[
%x[n]= A\theta+\eta[n] \qquad n=0, \ldots, N-1
%\]
%where $A$ is a known parameter and $\eta[\cdot]$ 
%is a sequence of independent Gaussian random variables with mean value $\mu$ and %variance
%$\sigma^2$. Given the independency condition, the Fisher information is additive. When $N=1$, 
%we have 
(App.~A): 
\begin{align}
\begin{split}
I_1(\theta) 
& \coloneqq
\sum_{n=-\infty}^{\infty} 
\frac{1}{2\pi\sigma^2}
\frac{1}{
p(n\Delta;\theta)}
\left\{ 
	e^{-\frac{1}{2\sigma^2}
	\left(
		-\nicefrac{\Delta}{2}+n\Delta -\theta 
	\right)^2} 
\right.\\
& 
\left.
	- e^{-\frac{1}{2\sigma^2}
	\left(
		\nicefrac{\Delta}{2}+n\Delta -\theta
	\right)^2}
\right\}^2,
\label{fisher}
\end{split}
\end{align}
and where
\[
p(x;\theta) \coloneqq
\Phi\left(\frac{\nicefrac{\Delta}{2}+x-\theta}{\sigma}\right)
 -\Phi\left(\frac{-\nicefrac{\Delta}{2}+x-\theta}{\sigma}\right),
\]
where $\Phi(\cdot)$ is the cumulative distribution function of a zero--mean 
Gaussian random variable with unity variance.
As a consequence, in the class of unbiased estimators, (\ref{crlbb}) becomes:
\be
	CRLB(\theta) = \left. CRLB_{b}(\theta)\right|_{m_e(\theta) \equiv 0} = \frac{1}{NI_1(\theta)}.
\label{crlb}
\ee

\begin{figure}[t]
\vskip-2.8cm
\begin{center}
\includegraphics[scale=0.48]{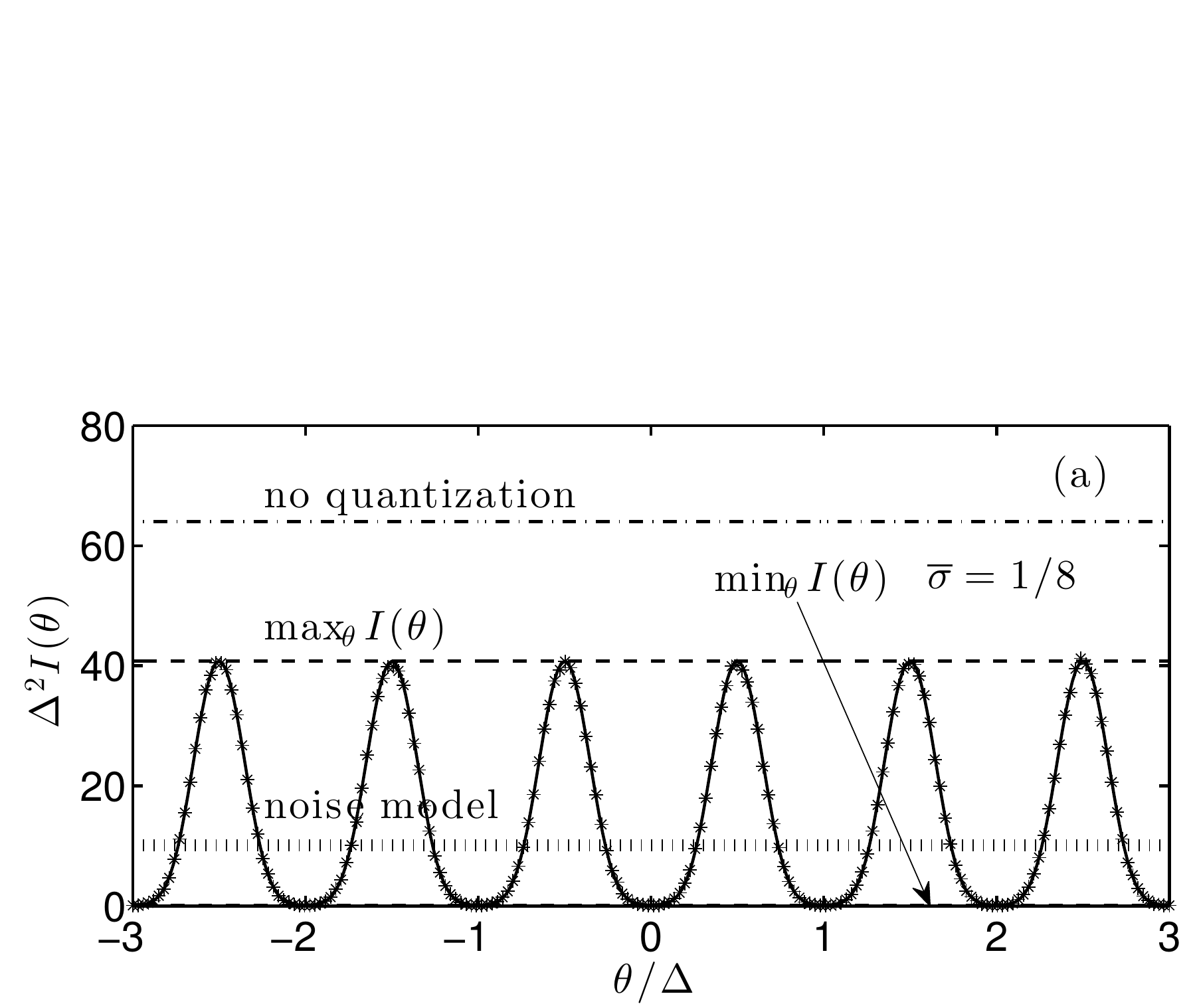}
\includegraphics[scale=0.48]{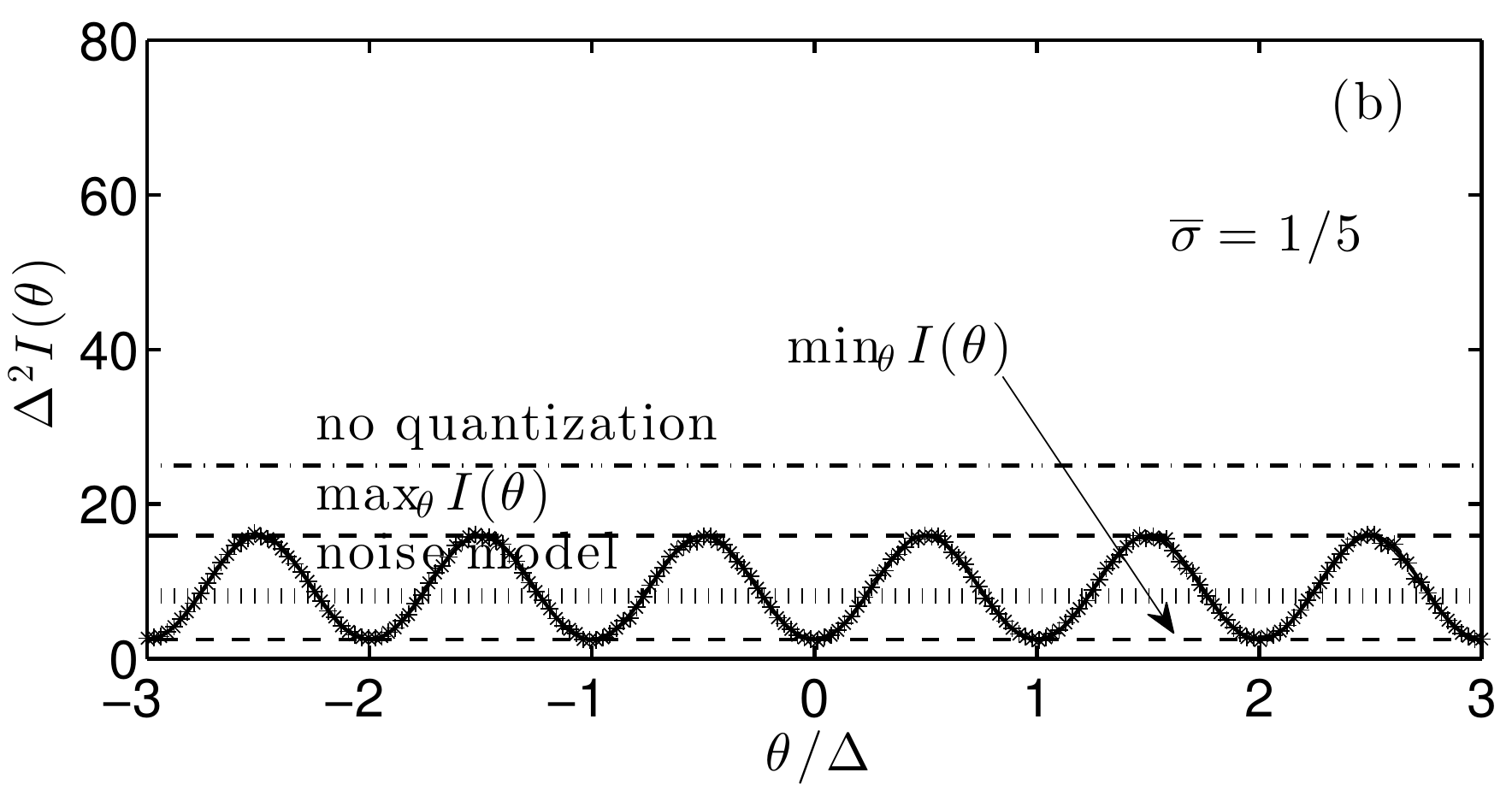}
\caption{Normalized Fisher information conveyed by a single sample of a quantized constant $\theta$ in Gaussian noise as a function of $\nicefrac{\theta}{\Delta}$ (solid line); the envelopes of maxima and minima with respect to $\theta$ are represented using dashed lines; also represented are the Fisher information associated to the application of the noise model of quantization  in absence of quantization. Stars represent simulation results based on $10^6$ Montecarlo runs. 
Figure (a) refers to $\overline{\sigma}=\nicefrac{1}{8}$ while figure (b) refers to $\overline{\sigma}=\nicefrac{1}{5}$. Plots are invariant with respect to $\Delta$ and thus with respect to the number of bits.\label{fig1}}
\end{center}
\end{figure}

An expression analogous to (\ref{crlb}) was derived in \cite{Gustafsson1}\cite{CarboneMoschitta}. In this paper this expression is further analyzed to show general properties of practical interest.
In App.~B it is shown that the Fisher information  $I_1(\theta)$ is a periodic function of $\theta$ with period $\Delta$ and that when $\overline{\sigma} <0.3 $ the maximum $I_M$ and minimum  values $I_m$ of $I_1(\theta)$, over $\theta$, are given respectively by:
\begin{equation}
	I_M \simeq \frac{2}{\pi \sigma^2}, \qquad \overline{\sigma} < 0.3
	\label{maxi}
\end{equation}
\begin{equation}
	I_m \simeq \frac{1}{\pi \sigma^2}\frac{e^{-\frac{1}{4\overline{\sigma}^2}}}{\Phi\left( \frac{3}{2\overline{\sigma}}\right)-\Phi\left( \frac{1}{2\overline{\sigma}}\right)}, \qquad \overline{\sigma} < 0.3.
	\label{mini}
\end{equation}
From (\ref{maxi}) and (\ref{mini}) it follows that maxima and minima in $\Delta^2 I_1(\theta)$
and the ratio $\rho(\overline{\sigma})\coloneqq \nicefrac{I_M}{I_m}$ only depend on the ratio $\overline{\sigma}=\sigma/\Delta$.
Expressions (\ref{fisher}), (\ref{maxi}) and (\ref{mini}) multiplied by $\Delta^2$ are plotted in Fig.~\ref{fig1} as a function of $\theta/\Delta$ for two different values of  $\overline{\sigma}=\nicefrac{1}{8}, \nicefrac{1}{5}$. Given that their behavior only depends on $\overline{\sigma}$, these curves scale with the number of bits.
In the same figure, the Fisher information applicable under the assumption 
of the noise model and when quantization is not applied, is also shown.
Results based on $10^6$ Montecarlo runs are plotted using stars to validate theoretical assumptions, being perfectly superimposed on theoretical expressions. 
Plots in Fig.~\ref{fig1}(a) show that when  $\overline{\sigma}$ is too small, the Fisher information vanishes for large intervals in the values $\theta/\Delta$, and the existence of a {\em consistent} estimator of $\theta$ becomes questionable \cite{KlaasenLenstra}. In this case quantization is destroying all available information and an interval estimator would  appear as a more suitable choice. 
In fact, observe that from (\ref{fisher}) we have
\be
	\lim_{\sigma \rightarrow 0^+} I_1(\theta) =  
	\sum_{n=-\infty}^{\infty}\delta\left( \theta - \left(n+\frac{1}{2}\right)\Delta \right), \qquad n \quad \mbox{integer},
\label{sigmazero}
\ee
where $\delta(\cdot)$ is the Dirac delta function. Expression (\ref{sigmazero}) shows that the Fisher information goes to infinity when $\sigma \rightarrow 0^+$ and the 
noiseless input value is positioned exactly in correspondence with a quantizer transition levels (which is of course not a realizable constraint in practice). 
In all other cases, when $\sigma \rightarrow 0^+$ the information vanishes.  
However, by artificially increasing this ratio, e.g. by {\em dithering},  the Fisher information can be made significantly larger than $0$, regardless of the value of $\theta$ \cite{CarbonePetri}.
In fact, Fig.~\ref{fig1}(b) shows that increasing  $\overline{\sigma}$ reduces the maxima in the Fisher information but also renders minima larger than $0$.  

\begin{figure}[t]
\begin{center}
\includegraphics[scale=0.48]{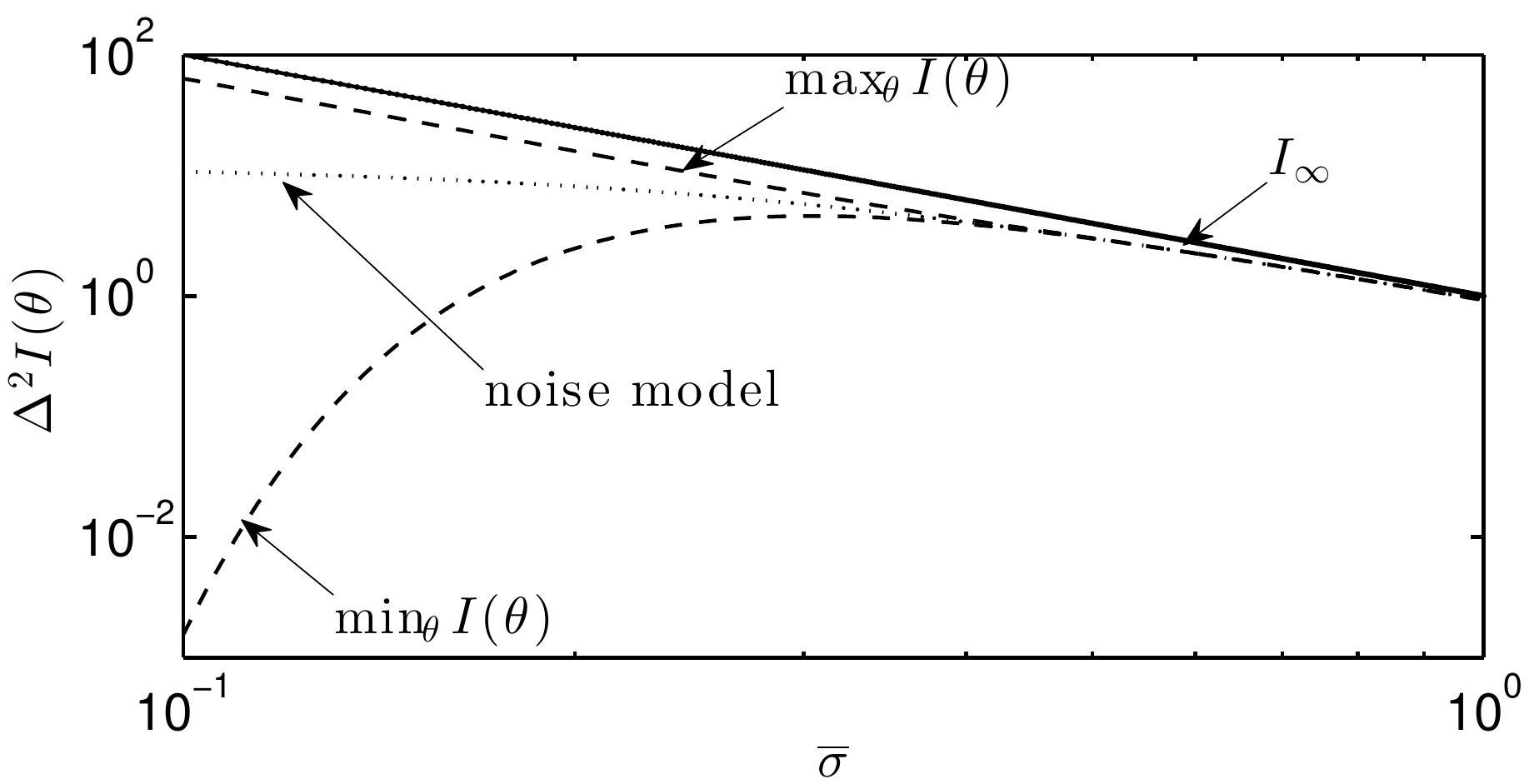}
\caption{Bilogarithmic plot of the normalized Fisher information conveyed by a single sample of a quantized constant $\theta$ in Gaussian noise as a function of $\overline{\sigma}\coloneqq\nicefrac{\sigma}{\Delta}$; maxima and minima with respect to $\theta$ are also shown together with the Fisher information associated to both the noise model of quantization and the absence of quantization. Plots are invariant with respect to $\Delta$ and thus with respect to the number of bits. \label{conv}}
\end{center}
\end{figure}

Consider that in the absence of quantization $I_1(\theta)=\nicefrac{1}{\sigma^2} \eqqcolon I_{\infty}$ \cite{Kay}. 
To show the behavior of the Fisher information as a function of
$\overline{\sigma}$,  (\ref{maxi}) and (\ref{mini})  are graphed in Fig.~\ref{conv} on a bilogarithmic scale, along with the Fisher information $I_q \coloneqq \nicefrac{1}{\left( \sigma^2+\frac{\Delta^2}{12}\right)}$, associated to the application of the noise model of quantization, and $I_{\infty}$. 
Observe that $\rho(0.3) \simeq 1.53$ and also that, for increasing values of $\overline{\sigma}$,
$\rho(\overline{\sigma}) \rightarrow 1$. Thus, for values of $\overline{\sigma}>\nicefrac{1}{3}$, oscillations in $I_1(\theta)$ with $\theta$ are greatly reduced and 
$I_q$ can be taken as a reasonable approximation of the Fisher information irrespective
of the value of $\theta$.

\begin{figure}[t]
\begin{center}
\includegraphics[scale=0.5]{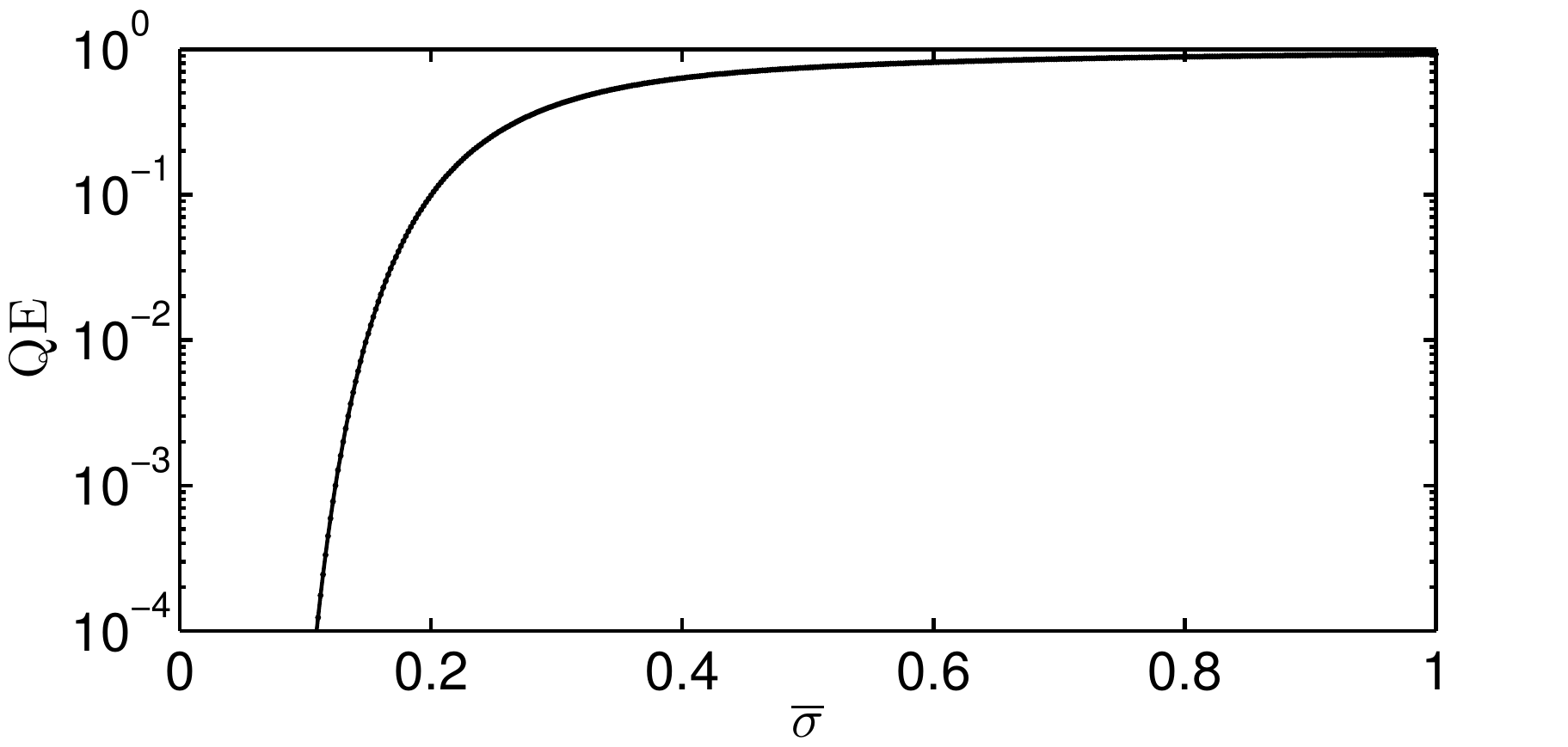}
\caption{{\em Quantization efficiency} (see text) as a function of $\overline{\sigma}$.
For small values of $\overline{\sigma}$ the worst case in the 
quantization of the noisy constant value introduces a significant loss in information that corresponds to a small value of QE. Conversely, when  $\overline{\sigma}\rightarrow 1$, 
QE tends
to $1$. In this case the effect of quantization is negligible.
The plot is invariant with respect to $\Delta$ and with respect to the number of bits.
\label{loss}}
\end{center}
\end{figure}

Derived expressions prove that quantization causes information loss. Accordingly,
{\em quantization efficiency} (QE) can be defined as follows:
\[
	QE \coloneqq \frac{\min_{\theta}I_1(\theta)}{I_{\infty}}.
\]
The behavior of QE, evaluated using (\ref{fisher}), is shown in Fig.~\ref{loss} as a function of $\overline{\sigma}$. When $\overline{\sigma}$
is close to $0$, quantization introduces a significant loss of information. Conversely when
$\overline{\sigma} \rightarrow 1$, the effect of quantization is negligible.
Thus, one possible critical value under which the loss is significant is again on the order of $\overline{\sigma}\simeq \nicefrac{1}{3}$, when this ratio becomes about equal to $0.5$. 
Consider that if (\ref{mini}) is used instead of (\ref{fisher}) a very good approximation is obtained of the plot in Fig.~\ref{loss}.

%Thus, for larger values of $\overline{\sigma}$ the noise model of quantization can reasonably 
%be applied and $I_1(\theta) \rightarrow \nicefrac{1}{\left( \sigma^2+\frac{\Delta^2}{12}\right)}$.

%For values of $\nicefrac{\sigma}{\Delta}$ larger than $0.3$. Consider that in absence of %quantization $I_1(\theta)=\nicefrac{A^2}{\sigma^2} \coloneqq I_{\infty}$ \cite{Kay}. 

\begin{figure}[t]
\begin{center}
\includegraphics[scale=0.5]{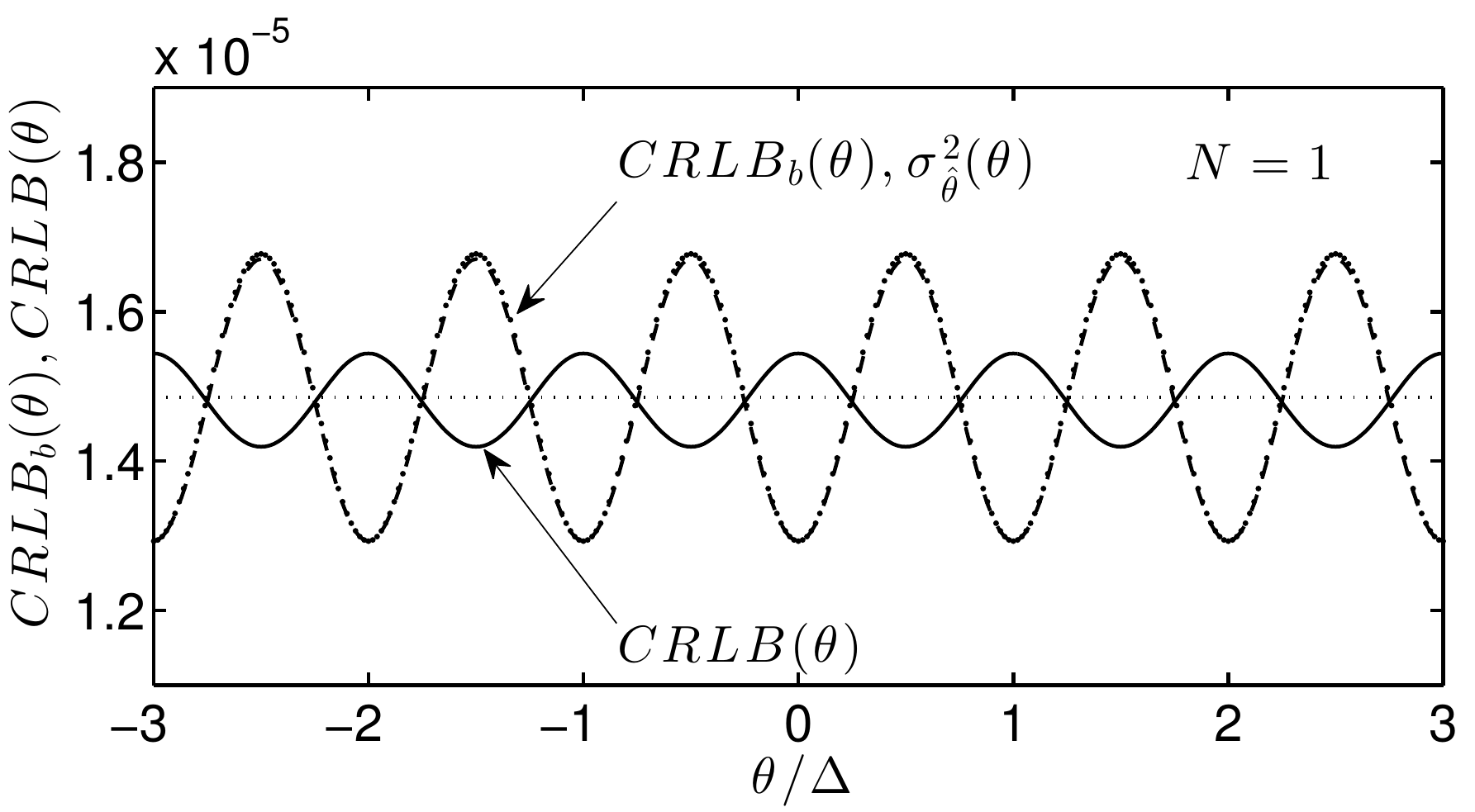}
\caption{Cramer--Rao lower bounds when estimating a quantized constant $\theta$ in noise: $CRLB(\theta)$ applies to unbiased estimators while $CRLB_b(\theta)$ is applicable to all estimators having bias given by (\ref{me_gau}). Shown is also the variance of $\hat{\theta}$ superimposed on $CRLB_b(\theta)$. \label{fig2}}
\end{center}
\end{figure}

\section{Performance of the Mean Value Estimator}
The statistical performance of (\ref{natest}) can now be assessed by 
comparing the variance of $\hat{\theta}$  with the CRLB. Since this estimator is biased (\ref{crlbb}) must be used. 
Thus, it is expected that 
%-
%- within the applicable approximations --
(\ref{ssyq}) will be larger than or equal to (\ref{crlbb}).
Expression (\ref{ssyq}) is plotted in Fig.~\ref{fig2} along with (\ref{crlbb})
and (\ref{crlb}) assuming  $N=1$.
Given the good match between $CRLB_b(\theta)$ and (\ref{ssyq}) it is argued that
the mean value of the sequence of random variables output by a uniform quantizer 
when the input is a constant plus Gaussian noise is an optimum estimator in the class
of estimators having the bias exactly given by (\ref{me_gau}).
Of course, in practice, the exact bias knowledge is not available because this requires also the exact knowledge  of the value of the mean.
One additional question is then if there are better estimators exhibiting lesser bias and possibly a variance closer to $CRLB(\theta)$. The next sections cover this topic.

\section{A moment--based estimator for a quantized constant in noise}
Provided that the arithmetic mean is optimal among estimators having an exactly known bias, 
in the considered estimation problem, the question remains whether there are other estimators with lower bias and this without assuming that the exact bias is known.
The goal of this section is to show the properties of alternative estimators 
characterized by better statistical properties than the simple arithmetic mean. 
Two approaches that appear as good candidates to achieve this goal
are the moment--based estimator and the maximum likelihood estimator, that are analyzed in this and in the next sections, respectively.

\subsection{Development of the estimator}
Estimation of parameters based on expressions of their statistical moments is a straightforward mechanism to obtain estimators or estimation procedures \cite{Kay}. Since theoretical 
expressions for the mean value and variance of a quantized constant are known, we can use both (\ref{me_gau}) and (\ref{s2xq_gau}) and find the value of $\theta$
that provides exactly these same values 
estimated by the usual sampling estimators of the mean value and of the quantizer output variance. Thus, at first ${\hat\theta}$
-- see (\ref{natest}) -- and 
\begin{equation}
	\hat{\sigma}^2_y = 
		\frac{1}{N-1}\sum_{n=0}^{N-1}
		\left( 
			y[n]-\hat{\theta}
		\right)^2,
\end{equation}
are evaluated using the available quantized samples.
Then, when (\ref{me_gau}) and (\ref{s2xq_gau}) hold true, we can 
define a moment--based estimator $\hat{\theta}_M$ of $\theta$ as follows \cite{Kay}:
\begin{equation}
	\hat{\theta}_M= 
	\left\{ 
		\theta    
		\Bigl|  
			\left[ \theta+{{m}}_e(\theta) \right]=\hat{\theta},
	\sigma_y^2(\theta)=
	\hat{\sigma}^2_y
	\right\},
	 \label{mom}
	\end{equation}
that is, that value of
$\theta$ that solves the two equations (\ref{me_gau}) and (\ref{s2xq_gau}) at the same time.
Consider that the solution of the system of nonlinear equations in (\ref{mom}) also requires the estimation of $\overline{\sigma}$ for  the evaluation of both (\ref{me_gau}) and (\ref{s2xq_gau}).
Thus, while
$\hat{\theta}$ does not require and does not use information about the noise standard deviation, $\hat{\theta}_M$ also estimates this parameter and it is thus expected to provide better estimates\footnote{if $\sigma$ is known, (\ref{mom}) simplifies because a single equation can be used to estimate $\theta$.}. 
Thus, once quantized values are used to obtain the two numerical estimates of the mean value and variance, the systems of two equations in (\ref{mom}) is solved numerically for $\theta$.   A similar approach was taken in \cite{Gustafsson2} where {\em dithering} noise was assumed satisfying the Schuchman condition \cite{Schuchman}. However, {\em dithering} is not always adopted
and satisfaction of the Schuchman condition can only approximately be fulfilled
given that quantizers used in practice are only nominally uniform.

Since (\ref{mom}) is biased, the 
Mean--Square--Error (MSE) defined as
\be
	mse_{\hat{\theta}^{\star}}(\theta) = E\left\{ (\theta-\hat{\theta}^{\star})^2\right\},
	%= m^2_e(\theta)+\sigma^2_e(\theta)
\label{mse}
\ee
is considered as a measure of effectiveness of the given estimator $\hat{\theta}^{\star}$.
In the case of  $\hat{\theta}^{\star}=\hat{\theta}$,
\be
	mse_{\hat{\theta}}(\theta) =  m^2_e(\theta)+\sigma^2_e(\theta).
\label{msemean}
\ee
Expression (\ref{msemean}) depends on $\theta$. Thus, its average  
$\overline{mse}$, that can be taken as an indicator of statistical performance, 
is evaluated over $M$ equally spaced points in the quantizer input interval.
Moreover, since both $m_{e}(\theta)$ and $\sigma_e^2(\theta)$ are periodic 
functions of $\theta$ with period $\Delta$, also 
({\ref{msemean}}) and $mse_{\hat{\theta}_M}(\theta)$ have the same property 
and thus it is sufficient to evaluate the average of $mse_{\hat{\theta}^{\star}}(\cdot)$
in the interval $\left( -\Delta/2, \Delta/2 \right)$, as follows:
\be
	\overline{mse}_{\hat{\theta}^{\star}} \coloneqq \frac{1}{M}\sum_{m=0}^{M-1}
	mse_{\hat{\theta}^{\star}}\left( -\frac{\Delta}{2}+m\frac{\Delta}{M}\right).
	\label{msea}
\ee

\subsection{Simulation results}
\begin{figure}[t]
\begin{center}
\vskip0.44cm
\includegraphics[scale=0.5]{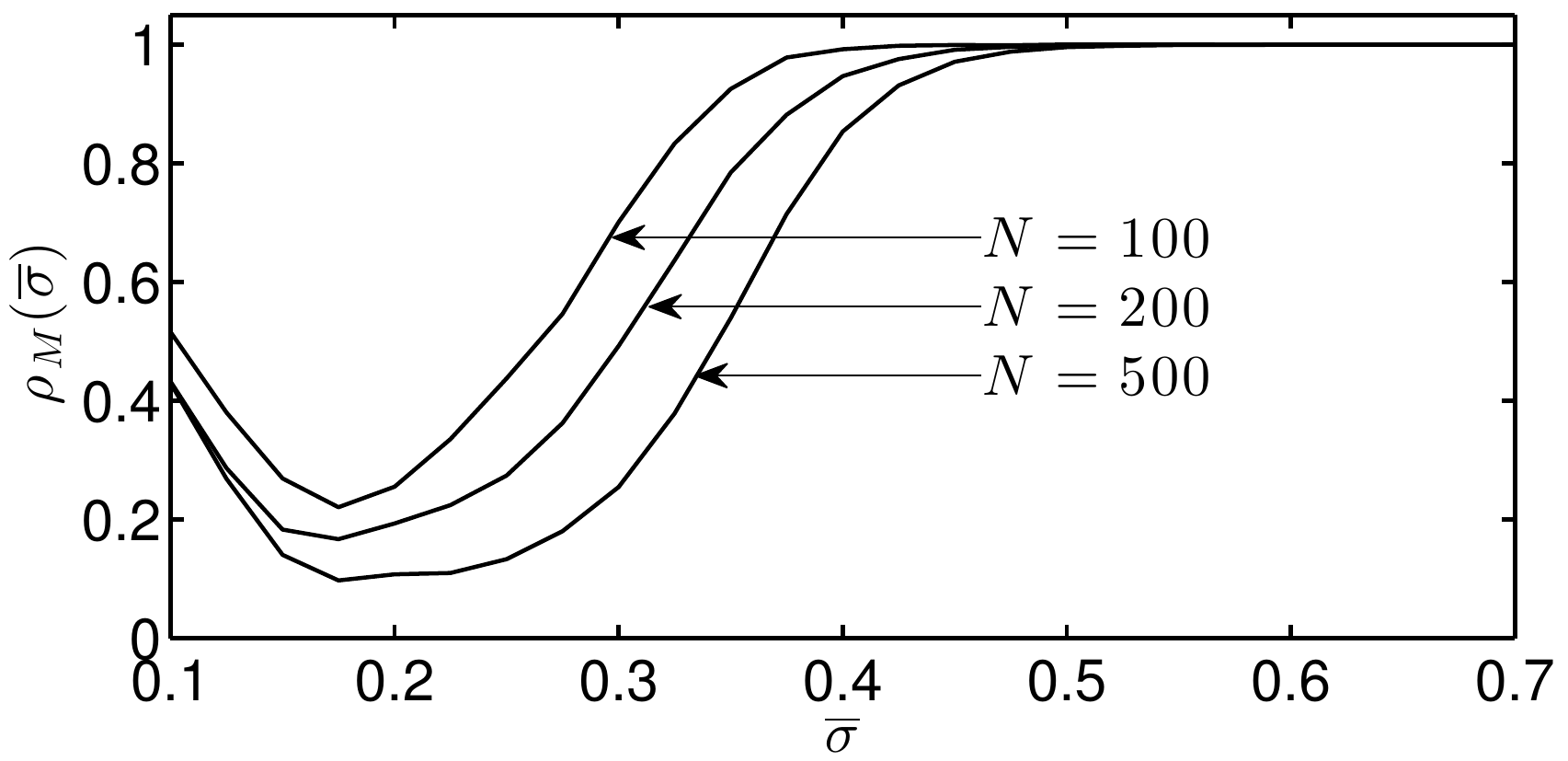}
\caption{Ratio between the average mean square error associated to the moment based estimator and to $\hat{\theta}$ as a function of $\overline{\sigma}=0.05\ldots 0.6$, with $\theta \in \left( -\frac{\Delta}{2}, \frac{\Delta}{2}\right)$. The moment based estimator outperforms $\hat{\theta}$ 
when $\overline{\sigma} \leq 0.4$, with efficiency increasing with $N$.\label{fig4}}
\end{center}
\end{figure}

To appreciate the statistical efficiency of the moment--based estimator
with respect to the arithmetic mean estimator, a 
Montecarlo--based analysis was performed using $R=200$ records and assuming $M=500$.
The ratio
\be
	\rho_M(\overline{\sigma}) \coloneqq \frac{\overline{mse}_{\hat{\theta}_M}}{\overline{mse}_{\hat{\theta}}},
\ee
between values of (\ref{msea}) evaluated
for both (\ref{natest}) and (\ref{mom}) was determined. Results obtained when assuming 
an $8$ bit uniform quantizer are plotted in Fig.~\ref{fig4}
as a function of $\overline{\sigma}$, for various values of $N$.
It can be observed that the moment--based estimator outperforms the 
arithmetic mean estimator when $\overline{\sigma} < 0.4$, that is when 
the {\em noise model} can not be applied. Observe also that while
(\ref{me_gau}) and (\ref{s2xq_gau}) hold for
$\overline{\sigma} > 0.3$, they still provide good results also for lower values of $\sigma$,
as they represent the first order terms in a Fourier series expansion \cite{CarbonePetri}. Higher accuracy is expected
by using a larger number of terms in the series, at the expense of a larger computational burden.

\section{A maximum likelihood estimator for a quantized constant in noise}
The problem of expressing the LF of 
(\ref{onebis})
was considered previously under different application scenarios \cite{Gendai}\cite{WangYinZhangZhao}\cite{Giaquinto}.
Here, the analysis is done to verify how the MLE provides 
better statistical performance over the arithmetic mean estimator, both when the quantizer is uniform and when nonlinearities affect the quantizer input--output characteristic. 

\subsection{Setting up the maximum likelihood estimator}
Given that a sequence of quantizer output values $y[n]$, $n=0, \ldots, N-1$
was observed, the LF is the probability of the quantizer interval around $\theta$, 
obtained by using the Gaussian noise assumption for $\eta[\cdot]$ in (\ref{onebis}) \cite{CarboneMoschitta}:
\be
l(\theta) 
	\coloneqq 
	\prod_{n=0}^{N-1}
	\left[ 
		\Phi
		\left(
			 \frac{y[n]+\frac{\Delta}{2}-\theta}{\sigma} 
		\right)-
 \Phi\left( \frac{y[n]-\frac{\Delta}{2}-\theta}{\sigma} \right) \right].
\label{likelihood}
\ee
Thus, the log--likelihood function becomes:
\begin{align}
\begin{split}
{\cal L}(\theta) \coloneqq \sum_{n=0}^{N-1}
&
\log \left[ \Phi\left( 
\frac{y[n]+\frac{\Delta}{2}-\theta}{\sigma} \right)- \right.\\
& \qquad \qquad \qquad \left.
\Phi\left( \frac{y[n]-\frac{\Delta}{2}-\theta}{\sigma} \right) \right].
\label{loglikelihood}
\end{split}
\end{align}
An alternate expression for the LF can be obtained by observing 
that the sequence $y[\cdot]$ contains a finite number of samples,
and that many of such samples may take the same value. 
Accordingly, define $N_i$ as the number of occurrences in the output sequence
observed in code--bin $i$, when acquiring $N$ samples.
Values for $N_i$ can be deduced by the histogram of the observed samples at the 
quantizer output and
%and ${\cal N}$
%\subset \{0,1,2,\ldots, L-1\}$ as the subset 
%as the set of integers $i$ for which $N_i>0$, 
we have
\[
	\sum_{i =-\infty}^{\infty}N_i = N.
\]
Thus, (\ref{likelihood}) becomes
\begin{equation}
l(\theta)  = 
	\prod_{i=-\infty}^{\infty}
	\left[ 
	\Phi
	\left( 
%		\frac{i\Delta+\Delta\frac{1-L}{2}	-\theta}{\sigma} \right)- 
		\frac{i\Delta+\frac{\Delta}{2}	-\theta}{\sigma} \right)- 
%	\Phi\left( \frac{(i-1)\Delta+\Delta\frac{1-L}{2}-\theta}{\sigma} \right) 
	\Phi\left( \frac{i\Delta-\frac{\Delta}{2}-\theta}{\sigma} \right) 
	\right]^{N_i}.
\label{likelihood2}	
\end{equation}
Finally, the MLE is determined by the value of $\theta$ and $\sigma$ that maximize any of the presented likelihood or log--likelihood functions (\ref{likelihood})--(\ref{likelihood2}). 
Observe that knowledge of the values of $N_i$ is sufficient to express (\ref{likelihood2})
as required by the hypotheses of  the {\em Neyman--Fisher} factorization criterion \cite{Kay}.
Thus, values of $N_i$, that is the histogram of quantized samples, 
represent {\em jointly sufficient statistics} 
for the estimation of $\theta$ in this estimation problem.
In addition, since $\hat{\theta}$ is a deterministic function of these 
jointly sufficient statistics, {\em Rao--Blackwellization} of this estimator would not lead 
to a better performing estimation procedure \cite{Kay}.

\subsection{Properties of the MLE}
In App.~C it is shown that the MSE $mse_{mle}(\theta)$ associated to the MLE is periodic with $\Delta$.
%if overloading effects can be neglected, that is if the quantizer is used within
%its no--overload input range, 
%the likelihood function is a periodic function of $\theta$
%with period $\Delta$. 
%Moreover by normalizing both $\theta$ and $\sigma$ by $\Delta$, one obtains:
%\begin{align}
%\begin{split}
%l(\theta)  = 
%	\prod_{i\in {\cal N}}
%	\left[ 
%	\Phi
%	\left( 
%		\frac{i+\frac{1-L}{2}	-\overline{\theta}}{\overline{\sigma}} \right)- 
%	\Phi\left( \frac{(i-1)+\frac{1-L}{2}-\overline{\theta}}{\overline{\sigma}} \right) 
%	\right]^{N_i}
%\label{likelihood3}	
%\end{split}
%\end{align}
%where $\overline{\theta}\coloneqq \frac{\theta}{\Delta}$.
%If overloading effects can be neglected,
%\begin{align}
%\begin{split}
%l(\theta)  = 
%	\prod_{i\in {\cal N}}
%	& 
%	\left[ 
%	\Phi
%	\left( 
%		\frac{i+\frac{1-L}{2}-\overline{\theta}}{\overline{\sigma}} \right)- 
%	\right.  \\
%	& \qquad \qquad
%	\left.
%	\Phi\left( \frac{(i-1)+\frac{1-L}{2}-\overline{\theta}}{\overline{\sigma}} \right) 
%	\right]^{N_i}
%\label{likelihood4}	
%\end{split}
%\end{align}
%
Thus, the evaluation of (\ref{mse}) for values of $\theta \in \left(-\frac{\Delta}{2},\frac{\Delta}{2}\right)$ is sufficient to characterize the statistical behavior of the MLE. 
As stated in \cite{Gustafsson1}, the MLE
suffers from the {\em curse of dimensionality}. However, it must be observed that 
often in practical situations, when low-- to medium--resolution ADCs are used, $\sigma$ is a fraction of  $\Delta$. Even if $\sigma$ exceeds $\Delta$, 
$N_i>0$ for 
few values of $i$. 
Thus, the numerical evaluation of (\ref{likelihood2}) can be confined to the product of those factors for which $N_i>0$. Moreover, since $\Phi(\cdot)$ significantly differs from $0$ and $1$ only in a limited interval of its argument, the 
numerical search for the value of $\theta$ and $\sigma$ that maximize (\ref{likelihood2})
can be confined to these intervals.  
%This limit significantly the cardinality of the set of integers values for which $N_i>0$ and thus 
%the amplitude of the interval 
%in which the maximum of the MLE must be sought for. 
%This search can be performed by 
%numerical computation. 
This search will produce the value of $\theta$ and $\sigma$ maximizing both
(\ref{likelihood}) and (\ref{loglikelihood}), for any sequence of $y[\cdot]$.
As in any numerical evaluation of maxima, the risk of obtaining local and not global maximizers
is to be considered. To mitigate this risk, a suitable tuning of the magnitude of the updating parameter in the numerical algorithm must be performed.
Observe also that when $\sigma \rightarrow 0^{+}$ 
the Fisher information tends to vanish for large intervals of $\theta$, as shown in Fig.~\ref{fig1}(b), and the maximum of the LF tends not to be unique.

\begin{figure}[t]
\vskip0.3cm
\begin{center}
\includegraphics[scale=0.5]{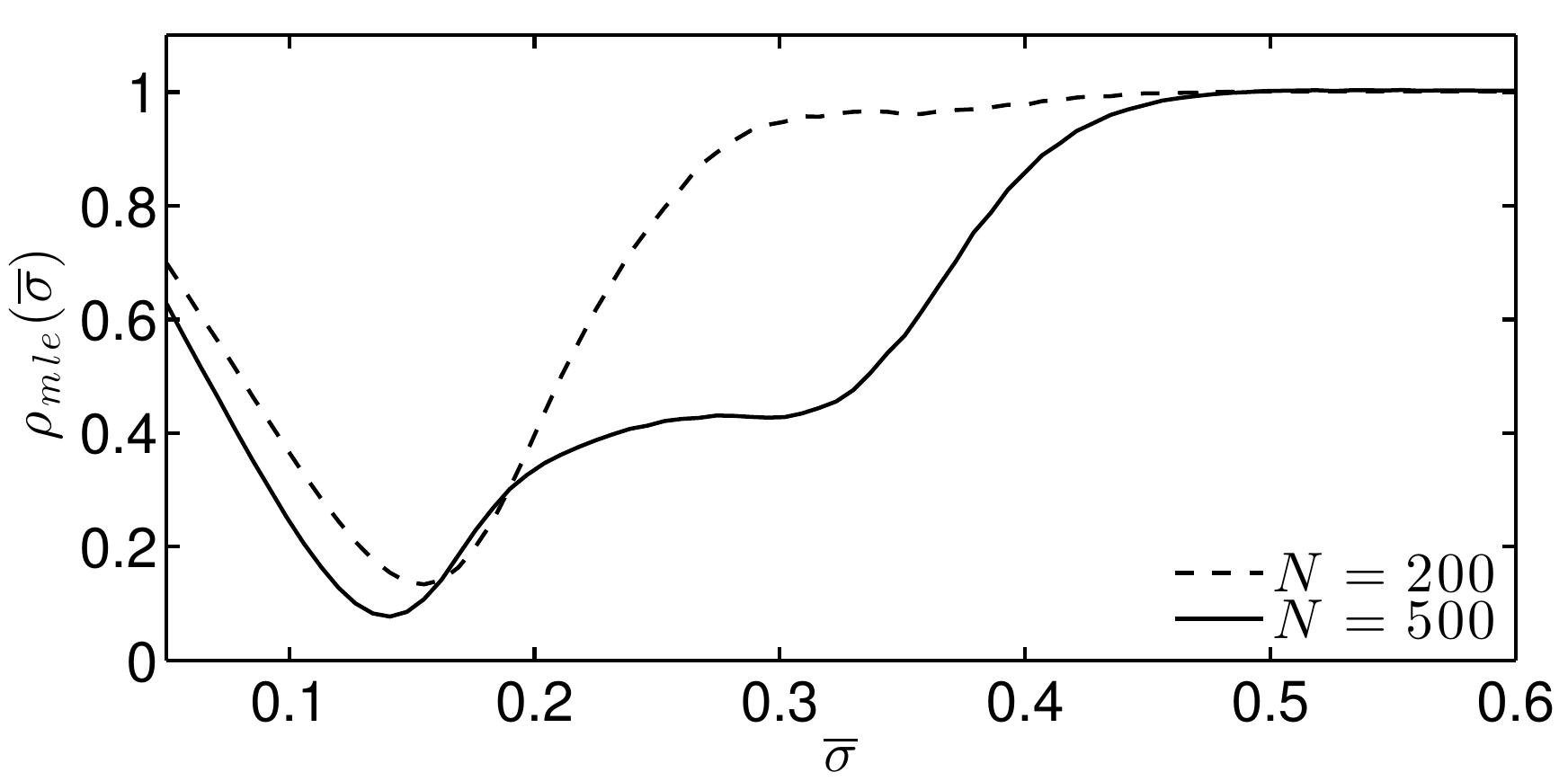}
\caption{Ratio between the average mean square error associated to the MLE and to $\hat{\theta}$ as a function of $\overline{\sigma}=0.05\ldots 0.6$, $\theta \in \left( -\frac{\Delta}{2}, \frac{\Delta}{2}\right)$. The MLE outperforms $\hat{\theta}$ 
when $\overline{\sigma} \leq 0.4$, with efficacy increasing with $N$. \label{fig5}}
\end{center}
\end{figure}

\subsection{Simulation results}
\begin{figure}[t]
\vskip-3cm
\begin{center}
\includegraphics[scale=0.5]{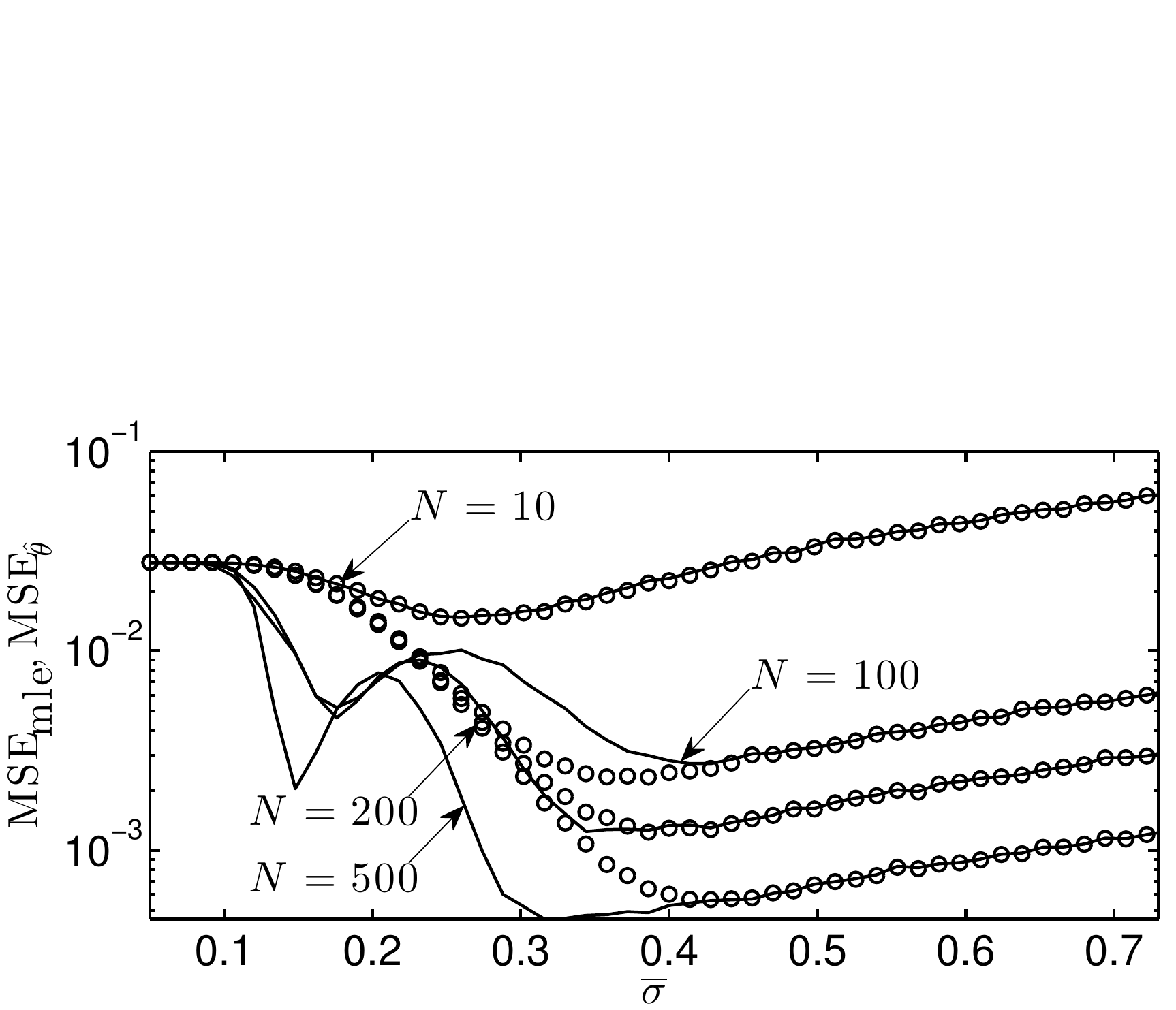}
\caption{Threshold effect of the MLE: Values of MSE associated to the arithmetic mean $\hat{\theta}$ (circles) and to
the MLE (solid line) assuming a constant input $\theta=\frac{\Delta}{6}$, as a function of $\overline{\sigma}=0.05\ldots0.6$, for several values of $N$. The MLE outperforms $\hat{\theta}$ uniformly only when $N=500$. \label{threshold}}
\end{center}
\end{figure}

To appreciate the gain in statistical efficiency obtained by using 
the MLE, $\rho_{MLE}(\overline{\sigma})$ defined as:
\[
	\rho_{mle}(\overline{\sigma}) \coloneqq \frac{\overline{mse}_{mle}}{\overline{mse}_{\hat{\theta}}}
\]
 was determined by a Montecarlo approach based on $2000$ records of $N=200, 500$
 values of a quantized constant in noise. To obtain results comparable to those obtained using the moment--based estimator, both $\theta$ and $\sigma$ were assumed unknown.
Then, the maximum in (\ref{likelihood2}) with respect to
 both $\theta$ and $\sigma$ was determined by means of a Nelder--Mead algorithm coded in C, initialized with the values of $\theta$ and $\sigma$ provided by sample estimators 
 \cite{Nelder}. Results are shown in Fig.~\ref{fig5}. As in the case of the moment--based estimator
 the MLE also outperforms the arithmetic mean estimator when $\overline{\sigma}<0.4$. 
 The fact that the estimator based on $N=500$ does not uniformly outperform the one 
 based on $N=200$ is due to the fact that this MLE appears to be affected by a {\em threshold} effect,
 when used in {\em nonasymptotic} conditions. This phenomenon was observed before
 when dealing with other estimation problems and is described, e.g., in \cite{ForsterLarzabalBoyer}.
In fact, for a given value of the DC input signal and of $\overline{\sigma}$ there is a minimum value of $N$ beyond which MLE has superior performance than the arithmetic mean estimator.
Below this value, this might not be true for all possible values of the problem parameters.
To show the nature of this problem consider 
the values of the MSE associated to $\hat{\theta}$ (circles) and $\hat{\theta}_M$ (solid line)
shown in Fig.~\ref{threshold} as a function of $\overline{\sigma}$, when assuming  $\theta=\frac{\Delta}{6}$.
While the two curves overlap when $N=10$, $\hat{\theta}$ outperforms  
$\hat{\theta}_M$ when $0.25 < \overline{\sigma} < 0.4$ and $N=100$.
The MLE uniformly outperforms the arithmetic mean estimator in this figure
only when $N=500$. The two estimators performs identically when $\sigma$ exceeds a
threshold that depends on $N$. 

\subsection{Effect of quantizer nonlinearity }
Practical quantizers embedded in commercial ADCs do not exhibit quantization thresholds
exactly positioned at integer or fractional parts of quantization steps. 
Instead, the position $T_i$ of the actual transition levels as defined, for instance, 
in \cite{IEEE1241}, differs from the corresponding nominal value $T_{i,nom}$.
To characterize this phenomenon integral and differential nonlinearities can be 
defined as
\begin{equation}
	INL_i \coloneqq \frac{T_i-T_{i,nom}}{\Delta}  
	\qquad 
	DNL_i \coloneqq \frac{T_i-T_{i-1}-\Delta}{\Delta}
\label{inldnl}
\end{equation}
when gain and offset errors are ignored \cite{IEEE1241}.
$INL_i$ captures the deviation of the actual from the nominal $i$--th transition levels. 
$DNL_i$ measures the difference between the actual and the nominal 
widths of the $i$--th quantization bin. 
If the absolute value of $INL_i$ is lower than $\nicefrac{1}{2}$, 
missing codes are avoided, and the monotonicity of the quantizer 
characteristic is maintained \cite{IEEE1241}.
At the same time, commercial ADCs are often designed to guarantee $DNL_i$ to be at least 
lower than $2$.
In both cases the LF in (\ref{likelihood2}) is likely to be wrong. 
In general it can be 
rewritten as:
\be
\prod_{i=-\infty}^{\infty}\left[ \Phi\left( \frac{T_{i+1}-\theta}{\sigma} \right)-
\Phi\left( \frac{T_i-\theta}{\sigma} \right) \right]^{N_i}.
\label{likelihood3}	
\ee

\begin{figure}[t]
\vskip1.5mm
\begin{center}
\includegraphics[scale=0.5]{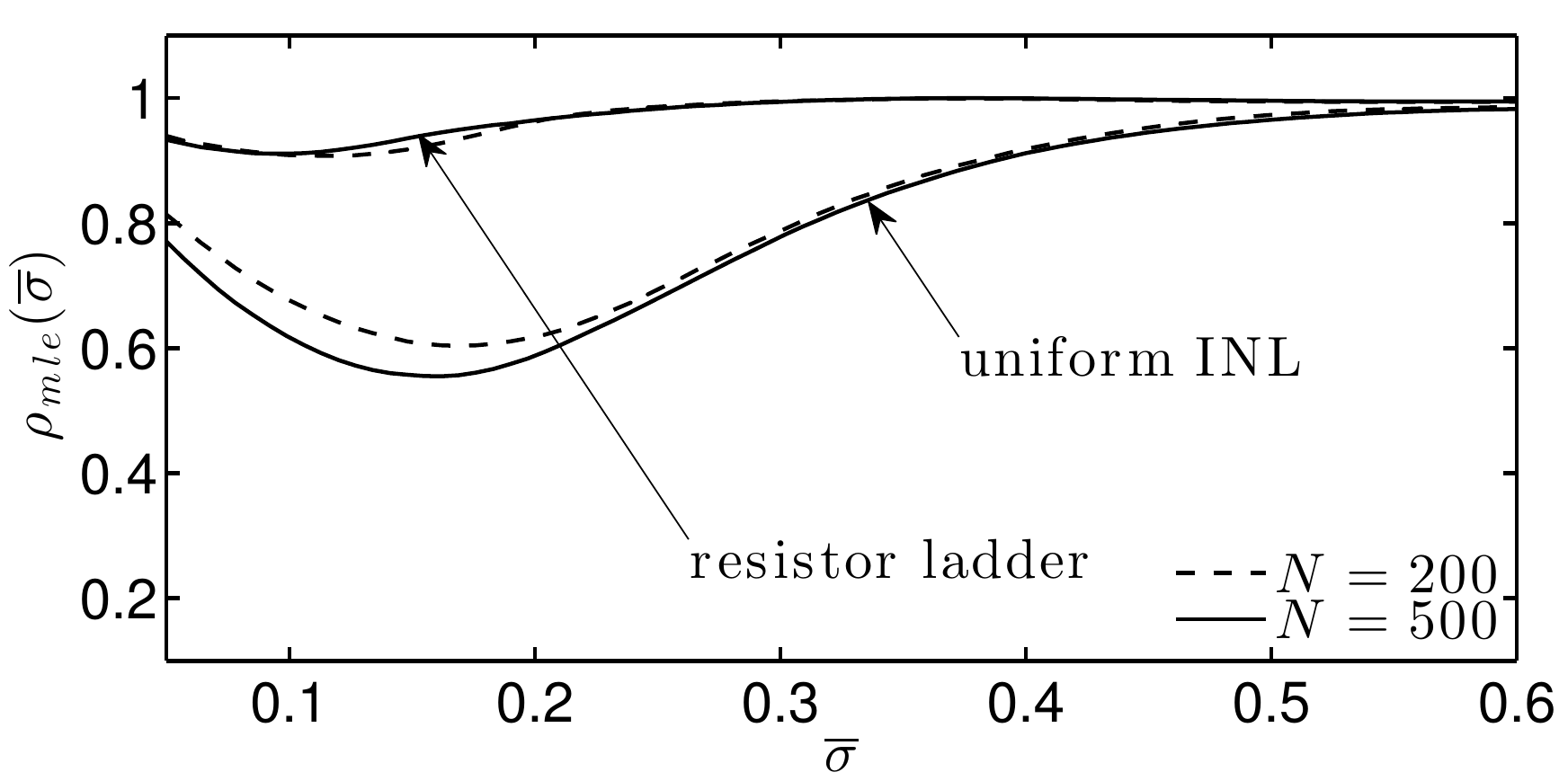}
\caption{Quantizer with nonlinearity uniformly distributed in $\left( -\frac{\Delta}{3}, \frac{\Delta}{3}\right)$,  and nonlinearity due to a resistor--ladder with resistances distributed as Gaussian random variables with mean $1000$\;$\Omega$ and standard deviation $150$\;$\Omega$. Ratio between average mean square errors associated to the MLE and to $\hat{\theta}$ as a function of $\overline{\sigma}=0.05\ldots 0.6$ and when $\theta \in \left( -\frac{25\Delta}{2}, \frac{25\Delta}{2}\right)$, assuming $N=200, 500$ and $R=2000$.   \label{nonlinear}}
\end{center}
\end{figure} 

%where $T_i$ represent the ADC transition levels, as defined for instance in 
%$\cite{IEEE1241}$. 
\noindent
If the sequence of $T_i$ is unknown two approaches can be taken:
either this fact is ignored and (\ref{likelihood2}) is used instead of (\ref{likelihood3}),
or the transition levels are treated as {\em nuisance} parameters, estimated as suggested for instance in \cite{IEEE1241}, and used to provide an approximated expression for the MLE\footnote{A third approach might be that of writing a LF to be maximized according to the values of $T_i$, $\theta$ and $\sigma$, with a large increase in computational complexity, being large the number of transition levels in ADCs.} in (\ref{likelihood3}). Clearly the resulted estimator will not be, in both cases, the true MLE, but it is interesting and useful from a practical viewpoint to understand if
this approach is  robust in this respect.
To appreciate the effect of integral and differential nonlinearities, 
$\rho_{mle}({\overline{\sigma}})$ is graphed
in Fig.~\ref{nonlinear}. It was evaluated using (\ref{likelihood}) as the LF.
Two relevant cases are considered. In the first case  
INL values are assumed to be uniformly distributed in the interval
$\left( -\nicefrac{1}{3}, \nicefrac{1}{3}\right)$ assuming $N=200, 500$. 
In the second case a resistor--ladder based ADC is simulated, assuming resistances
drawn from a Gaussian distribution with coefficient  of variation of $15\%$ \cite{resistorladder}.
This architecture guarantees monotonicity of the quantizer while allowing for rather large
values of INL. The simulated ADC is characterized by maximum magnitudes of DNL and INL, respectively bounded by $0.5$ and $2$. 
Observe that, while (\ref{mse}) is periodic with $\Delta$ when the quantizer is uniform, the MSE in the presence of nonlinearities is no longer periodic.
Thus, its evaluation requires spanning of $\theta$ of several quantization bins. 
In Fig.~\ref{nonlinear}, $R=2000$ records of samples are processed with $\theta$
spanning the interval $\pm \nicefrac{(25\Delta)}{2}$ instead of $\pm \nicefrac{\Delta}{2}$.
Even though uncertainty is added in the system, the MLE still outperforms the mean value estimator in a large meaningful interval of values for $\overline{\sigma}$ in the case of moderate amount of integral nonlinearities. Conversely, in the case of the resistor--ladder based ADC, the amount of INL and DNL is rather large and the increase in statistical efficiency obtained using
(\ref{likelihood}) is rather low. 
It may be concluded that the simplified usage of the MLE based on (\ref{likelihood}) is enough robust to accommodate for moderate amounts of INL in nominally uniform quantizers. 

\subsection{Experimental results}
To validate results presented in this paper, data were collected using a commercial $12$--bit USB data acquisition board (DAQ), sampling at a $10$ ksample/s rate.
A commercial PC--controlled synthesized signal generator was used to generate a constant input affected by Gaussian noise. Similarly, a commercial PC-controlled $6$\textonehalf--digit digital multimeter was employed to measure -- as a reference -- {\em a true quantity value}  $V_{\theta}$ of the measurand \cite{VIM}.
The experimental setup was used to measure $500$ samples uniformly distributed in the $\pm 4.5\Delta$ of the DAQ input range about $0$ V.  The used configuration for the DAQ provided $\Delta=5.08$~mV.  Moreover, DAQ data were compensated for the presence of an offset  component, estimated using the collected data.
Estimators based on the arithmetic mean, on the moments and on the maximum--likelihood approach were used to find the average estimation error $m_{\epsilon}(\overline{\theta})$ for each value of $\theta$, based on $10$ records. The estimation error was determined as the difference between the values provided by the three estimators and $V_{\theta}$. 
Estimators did not take into account the simulated INL and assumed the quantizer to be the ideal one described by (\ref{quant}).
%were used ignoring the contributions of the INL, considered as a
%{\em nuisance} parameter, and thus assuming 
%the only known value of $\Delta$. 
Moreover, both the moment--based estimator and the MLE 
were used without prior knowledge of the noise standard deviation.

\begin{figure}[t]
\vskip1.5mm
\begin{center}
\includegraphics[scale=0.5]{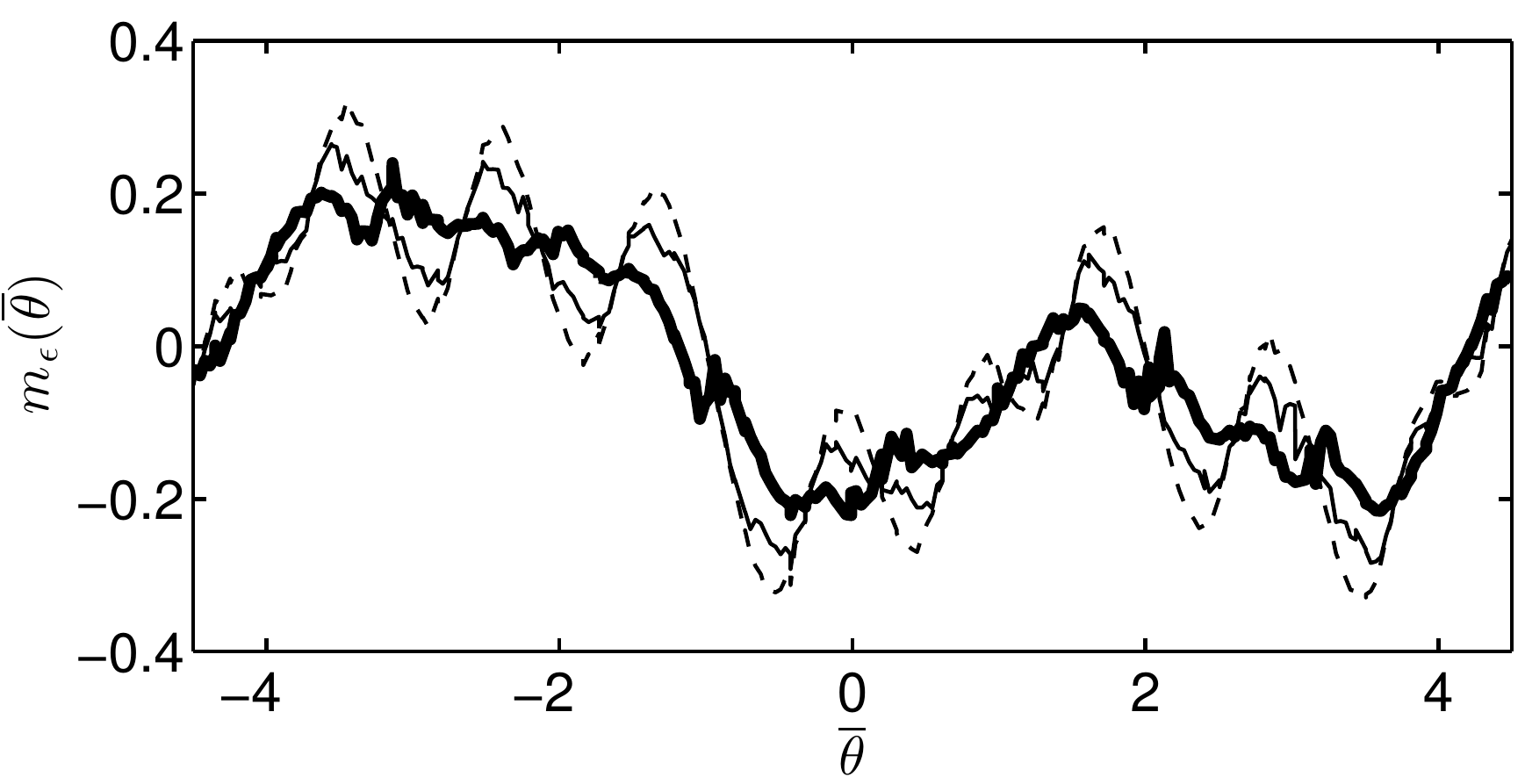}
\caption{Experimental results obtained using a commercial $12$--bit data acquisition board having INL: 
mean value of the estimation error expressed in fractions of $\Delta$ as a function of $\overline{\theta}$, in the case of the simple arithmetic mean--based (dashed line), of the moment--based estimator (solid line) and of the MLE (bold solid line). For each input value, $10$ records are used to evaluate averages, in all cases.
\label{nonlinearexp}}
\end{center}
\end{figure}

Results, normalized to $\Delta$, are shown in Fig.~\ref{nonlinearexp} using a bold solid line (MLE), a solid line (moment--based estimator) and a dashed line (arithmetic mean). 
The behavior of $m_{\epsilon}(\overline{\theta})$ in the case of the arithmetic mean estimator (dashed line) is not periodic, but it rather shows a superimposed low--frequency component (LFC) \cite{Handel}. Clearly, the used DAQ is not a uniform one, and it is affected by INL.
The dashed line in Fig.~\ref{nonlinearexp}
also highlights the {\em dithering} effect of the noise superimposed on the DC value that smooths the sharp variations typical of the quantization error input--output characteristic. 
The MLE provides an overall smaller error than that associated to the arithmetic mean, with
an experimental $\rho_{MLE}(\overline{\sigma}) \simeq 0.65$, with an estimated $\overline{\sigma} \simeq 0.21$. This value is compatible with results shown in Fig.~\ref{nonlinear}. 
Observe also that the MLE approximately fits the LFC of the INL and smooths its high--frequency variations.  The moment--based estimator has a performance in between the other two estimators and its usage provides $\rho_{M}(\overline{\sigma}) \simeq 0.79$.

%The application of the moment--based estimator provided no improvements over the arithmetic %mean estimator. This is due to the non--uniform behavior of the quantizer in the DAQ. The %moment--based estimator in fact tries to fit the sinusoid--like error shape given by (\ref{me_gau}) to %the measured one. Because of the LFC, improvements by this fitting, when the algorithm is %initialized with sample mean and sample standard deviations,  is not achieved.  

\subsection{Discussion of results}
Based on the results presented 
in this paper,
when DC values affected by Gaussian noise are quantized, it can be stated that:
\begin{itemize}
\item When the noise standard deviation is larger than or comparable to about half the width of the quantization step, the noise model of quantization can reasonably be applied and the estimation bias of the arithmetic mean estimator can largely be tolerated, as it vanishes quickly when the noise intensity is increased. This is evident from the analysis of the behavior of the Fisher information and from the statistical performance of the arithmetic mean estimator;
\item When the noise standard deviation is lower than about half the width of the quantization step, significant bias may affect the arithmetic mean estimator. Two other estimators were presented in this paper: both of them provide a better mean MSE than the arithmetic mean estimator, especially for low values of $\overline{\sigma}$. 
While the MLE 
is known to have desirable asymptotic properties (such as asymptotic efficiency) it does not  uniformly  exhibit superior performance over the moment--based estimator 
when a finite number of samples is collected. Moreover, numerical simulations show that it is not insensitive to the values assigned to the parameters of 
the used Nelder--Mead algorithm that must be tuned to reduce the risk of incurring in
local maxima. Even if also the moment--based estimator is evaluated numerically, tuning of the corresponding parameters in the numerical algorithm is less critical. However,
its asymptotic behavior is not predictable as in the case of the MLE and  it is less robust in the presence of quantizer nonlinearity. Thus, when a small number of samples is collected and the quantizer is uniform,  the moment--based estimator provides a reasonable improvement  
in statistical performance over the usage of the simple arithmetic mean. On the other hand when the number of samples is large or nonlinearites affect the quantizer, the MLE may represent a better choice.
\end{itemize}

%We have:
%\begin{align}
%\sum_{n=a}^{b}f(n) & =
%\int_{a}^bf(x)dx -B_1\left(f(b)+f(a) \right)\\
%& + \sum_{k=1}^{p}\frac{B_{2k}}{(2k)!}\left( f^{(2k-1)}(b)-f^{(2k-1)}(a)\right)+R
%\end{align}
%where
%\[
%R = \int_a^bf^{(2p+1)}(x)\frac{P_{2p}(x)}{(2p)!}dx
%\]

\section{Conclusion}
In this paper we considered the estimation of a constant value affected by Gaussian noise and quantization. No simplifying assumptions are made on the effects of quantization whose 
impact on the available information is first determined by expressing the Cram\'er--Rao lower bound and its maxima and minima, as the constant varies.
The properties of the arithmetic average, the moment--based and the maximum--likelihood 
estimators are analyzed also when nonlinearities cause the transition levels in the quantizer to be non uniformly distributed.
Both simulation and experimental results show that improvement in accuracy over the simple average are obtained especially when the noise standard deviation is comparable to the quantization step. It is also shown that under non--asymtptotic conditions, the maximum likelihood estimator is affected by a threshold effect and that it does not necessarily outperform the simple average estimator.
Overall, an accuracy improvement is obtained at a moderate increase in computational complexity.

The provided  expressions and procedures can profitably be used in numerical instrumentation and 
in system identification to obtain better estimates when using quantized data. While the analyzed signal model is based on a constant input value it can be extended to time--varying parametric deterministic signals by using the same analytical techniques.
This is done for instance in \cite{Kollar1}\cite{Kollar2}, where the meaningful problem of estimating the parameters
of a sinewave is considered in the context of ADC testing.

\appendix
\section{Evaluation of the Cram\'er--Rao\\ lower bound}
%If
%\[
%E\left[ \frac{\partial \ln p(x;\theta)}{\partial \theta } \right] = 0 \quad \forall \theta
%\]
%then
%\[
%\mbox{var}(\theta) \ge \frac{1}{-E\left[ \frac{\partial \ln^2 p(x;\theta)}{\partial \theta^2 } \right]} =
%\frac{1}{E\left[\left(  \frac{\partial \ln p(x;\theta)}{\partial \theta } \right)^2 \right]}
%\]
A similar approach to that taken in \cite{Gustafsson2} is taken here to find an expression of the 
Cram\'er--Rao lower bound.
If $X$ is a random variable, 
\[
Y\coloneqq Q(X)
\]
where $Q(\cdot)$ is a mid tread uniform quantizer, the probability density function of
$Y$ is \cite{KollarBook}:
\[
f_Y(y) = \left\{ (f_{\Delta} \ast f_X)(y) \right\} c_y(y)
\]
where $f_{\Delta}(\cdot)$ is the probability density function of a random variable uniform in $(-\nicefrac{\Delta}{2},\nicefrac{\Delta}{2})$ and 
\[
c_y(y) = \Delta\sum_{n=-\infty}^{\infty}\delta(y-n\Delta)
\]
is a pulse train.
Then
\[
f_X(x) = \frac{1}{\sqrt{2\pi}\sigma}e^{ 
-\frac{1}{2\sigma^2}\left( x-\theta\right)^2}
\]
The convolution with $f_{\Delta}(\cdot)$ is equivalent to low--pass filtering
the Probability Density Function (PDF) of $X$ with a filter having a {\em sinc} behavior in frequency. So, it is expected that
the convolved PDF be a smoothed version of the original PDF.
Thus,
\begin{align}
\begin{split}
f_Y(y) & =c_y(y) \int_{-\infty}^{\infty}f_{\Delta}(x)f_{X}(y-x)dx  \\
&  =c_y(y)\int_{-\nicefrac{\Delta}{2}}^{\nicefrac{\Delta}{2}}\frac{1}{\Delta} f_{X}(y-x)dx \\
& = c_y(y)\int_{-\nicefrac{\Delta}{2}}^{\nicefrac{\Delta}{2}}\frac{1}{\Delta}
\frac{1}{\sqrt{2\pi}\sigma}e^{ 
-\frac{1}{2\sigma^2}\left( y-x-\theta\right)^2}dx \\
& = \frac{1}{\Delta}c_y(y)\left\{ 
\Phi\left(\frac{\nicefrac{\Delta}{2}+y-\theta}{\sigma}\right)\right.\\
 & -\left.\Phi\left(\frac{-\nicefrac{\Delta}{2}+y-\theta}{\sigma}\right)
\right\} \\
&  = \sum_{n=-\infty}^{\infty}\delta(y-n\Delta)p(y;\theta) 
\end{split}
\end{align}
where 
%\begin{align}
%\begin{split}
\be
p(y;\theta)  = 
\Phi\left(\frac{\nicefrac{\Delta}{2}+y-\theta}{\sigma}\right)-\Phi\left(\frac{-\nicefrac{\Delta}{2}+y-\theta}{\sigma}\right)
\label{py}
%\end{split}
%\end{align}
\ee
and $\Phi(y)= \int_{-\infty}^y \frac{1}{\sqrt{2\pi}}e^{ 
-\frac{x^2}{2}}dx$.
Consider the Fisher information 
\begin{align}
\begin{split}
I_1(\theta)  & = 
E\left[\left(  \frac{\partial \ln f_Y(y;\theta)}{\partial \theta } \right)^2 \right]
= E\left[\frac{1}{f_Y^2(y;\theta)}  \left(  \frac{\partial f_Y(y;\theta)}{\partial \theta } \right)^2 \right].
\end{split}
\end{align}
Thus,
\begin{align}
\begin{split}
I_1(\theta) & =
\int_{-\infty}^{\infty} \frac{c^2_y(y)}{\Delta^2}\frac{1}{f_Y(y;\theta) }\frac{1}{2\pi \sigma^2}
\left\{
e^{-\frac{1}{2\sigma^2}
\left(-\nicefrac{\Delta}{2}+y -\theta \right)^2} \right.\\
& 
\left.
-
e^{-\frac{1}{2\sigma^2}
\left(\nicefrac{\Delta}{2}+y -\theta \right)^2}
\right\}^2 dy,
\end{split}
\end{align}
from which (\ref{fisher}) results.

%\appendix
\section{Properties of the Cram\'er--Rao\\ lower bound}
To prove the periodicity of the Fisher information with $\Delta$ 
consider that with $m$ as an integer value we have:
\begin{align}
I_1\left(\theta+m\Delta\right)  =
\sum_{n=-\infty}^{\infty} 
a_n\left(\theta+m\Delta\right),
\label{periodf}
\end{align}
where
%\begin{align}
%\begin{split}
\be
a_n(\theta) \coloneqq
%&
\frac{1}{2\pi\sigma^2}
\frac{ \left[
	e^{-\frac{1}{2\sigma^2}
	\left(
		-\nicefrac{\Delta}{2}+n\Delta -\theta
	\right)^2} 
%\right.
%\left.
	- e^{-\frac{1}{2\sigma^2}
	\left(
		\nicefrac{\Delta}{2}+n\Delta -\theta
	\right)^2}
\right]^2
}{
\Phi\left(\frac{\nicefrac{\Delta}{2}+n\Delta-\theta}{\sigma}\right)
 -\Phi\left(\frac{-\nicefrac{\Delta}{2}+n\Delta-\theta}{\sigma}\right)
}.
%\times
% \\
%& \hskip-0.35cm
\label{fisherapp}
%\end{split}
%\end{align}
\ee
From (\ref{fisherapp}) it follows that $a_{n}(\theta+m\Delta)=a_{(n-m)}(\theta)$. Thus, by changing the summation index in (\ref{periodf}) from $n$ to $k=n-m$, the periodicity property remains proved. It can also be proved that $I_1(\theta)$ is an even function of
$\theta$. Observe also that 
$I_1(\theta)=I_1(\Delta-\theta)$. Thus, if $\theta^{'}=\theta+\nicefrac{\Delta}{2}$ we have $I_1(\theta^{'})=I_1(-\theta^{'})$ and $I_1(\theta)$ is a symmetric function about 
the value $\theta=\Delta/2$. Because of this latter statement and since $I_1(\theta)$ is
periodic with period $\Delta$, to find its 
minimum and maximum, only the values
$\theta \in \left[0, \frac{\Delta}{2} \right]$ can be considered.

The derivative of $a_n(\theta)$ with respect to $\theta$, is zero when $\theta=n\Delta$.
Since $I_1(\theta)$ is periodic with period $\Delta$, $\theta=0$ minimizes each $a_n(\cdot)$ and thus the entire summation. Observe that the function at the numerator of $a_n(\cdot)$ is 
vanishing  faster than that at the denominator with increasing values of $n$, when $\theta=0$. Moreover, consider that each term of the type $\exp{\left( -\frac{\beta}{2\overline{\sigma}^2}\right)}$ can be neglected with respect to other terms in (\ref{fisherapp}) when $\beta >1$ and $\overline{\sigma} < 0.3 $.
Thus when $\overline{\sigma} < 0.3$,
the value of the exponentials
in $a_n(\theta)$ is appreciably larger than $0$ only when $n=-1,0,1$. 
Being $a_0(0)=0$, the minimum value of $I_1(\cdot)$ becomes
\begin{align}
\begin{split}
I_m = & \min_{\theta}I_1(\theta) =  I_1(0)  \simeq 
%\frac{A^2}{2\pi\sigma^2}
 a_{-1}(0) + a_1(0) \\
= &  
\frac{\left[ \frac{1}{\sqrt{2\pi}\sigma} \exp\left( 
{-\frac{9}{8}\frac{1}{\overline{\sigma}^2}}\right)-\frac{1}{\sqrt{2\pi}\sigma} \exp{\left( 
-\frac{1}{8\overline{\sigma}^2}\right)}\right]^2}
{\Phi\left(-\frac{1}{2\overline{\sigma}}\right)
 -\Phi\left(-\frac{3}{2\overline{\sigma}}\right)
}\\
+& \frac{\left[ \frac{1}{\sqrt{2\pi}\sigma} \exp{\left(-\frac{1}{8\overline{\sigma}^2}\right)}-\frac{1}{\sqrt{2\pi}\sigma} \exp{\left(
-\frac{9}{8}\frac{1}{\overline{\sigma}^2}\right)}\right]^2}
{\Phi\left(\frac{3}{2\overline{\sigma}}\right)
 -\Phi\left(\frac{1}{2\overline{\sigma}}\right)
} \\
\simeq &
\frac{2
\left[
\frac{1}{\sqrt{2\pi}\sigma} 
\exp{
	\left(
		-\frac{1}{8\overline{\sigma}^2}
	\right)
}
\right]^2
}
{\Phi\left(\frac{3}{2\overline{\sigma}}\right)
 -\Phi\left(\frac{1}{2\overline{\sigma}}\right)}, \qquad \overline{\sigma}<0.3 
\end{split}
\end{align}
from which (\ref{mini}) results.

To find the maximum of $I_1(\cdot)$ consider again that, since $I_1(\theta)$ is a periodic and even function of $\theta$ and it is symmetric about the value $\theta=\Delta/2$, the search for the maximum of 
$I_1(\theta)$ can be confined to the interval $\theta \in [0, \nicefrac{\Delta}{2}]$. 
When $\overline{\sigma}<0.3$, $a_n(\theta)$ is appreciably larger than $0$ only when $n=-1,0,1$. Observe also that
$a_{-1}(\theta)=a_0(\theta+\Delta)$ and $a_1(\theta)=a_0(\theta-\Delta)$ so that
\be
	I_1(\theta) \simeq a_0(\theta)+a_0\left(\theta-\Delta\right)+a_0\left(\theta+\Delta\right)
	 \qquad \overline{\sigma}<0.3 
\ee
By equating its derivative to $0$, we have
\be
	a^{'}_0(\theta)+a_0^{'}\left(\theta-\Delta\right)+a_0^{'}\left(\theta+\Delta\right)=0
\label{deriv}
\ee
where $a^{'}_0(\theta)$ is the derivative of $a_0(\theta)$. 
It can be verified by direct substitution that
the value $\theta=\nicefrac{\Delta}{2}$ renders  $a_0^{'}\left(\theta+\Delta\right)$
negligible with respect to the other two terms that sum to $0$.
% (\ref{deriv}) 
% -- the other value is $\theta=0$ --  and
Thus $\theta=\nicefrac{\Delta}{2}$ is the value that maximizes $I_1(\theta)$.
Then we have:
\begin{align}
\begin{split}
I_M & \simeq  I_1\left( \frac{\Delta}{2}\right) \simeq a_0\left( \frac{\Delta}{2}\right)+
a_1\left( \frac{\Delta}{2}\right)\\
& =
\frac{\left[ \frac{1}{\sqrt{2\pi}\sigma} \exp\left( 
{-\frac{1}{2\overline{\sigma}^2}}\right)-\frac{1}{\sqrt{2\pi}\sigma}\right]^2}
{0.5
 -\Phi\left(-\frac{1}{\overline{\sigma}}\right)
}+\\
& + \frac{\left[ \frac{1}{\sqrt{2\pi}\sigma}
 -\frac{1}{\sqrt{2\pi}\sigma} \exp{\left(
-\frac{1}{2\overline{\sigma}^2}\right)}\right]^2}
{\Phi\left(\frac{1}{\overline{\sigma}}\right)
 -0.5 }  \qquad \overline{\sigma}<0.3.
 \\
\end{split}
\end{align}
When $\overline{\sigma}<0.3$ the exponentials can be neglected, 
$\Phi\left(-\frac{1}{\overline{\sigma}}\right) \simeq 0$,
$\Phi\left(\frac{1}{\overline{\sigma}}\right) \simeq 1$ and
(\ref{maxi}) results.

\section{Properties of the likelihood function}
By observing that any real number $x$ can be written as $x=\lfloor x \rfloor +\langle x \rangle$,
where $\langle x \rangle$ represents the fractional part operator, from (\ref{likelihood2}) we have:
\begin{align}
\begin{split}
l(\theta)  = 
	\prod_{i=-\infty}^{\infty} &
	\left[ 
	\Phi
	\left( 
		\frac{i\Delta+\frac{\Delta}{2}	- \Delta\lfloor \frac{\theta}{\Delta} \rfloor-\Delta\langle \frac{\theta}{\Delta} \rangle}{\sigma} \right)- \right. \\
		& \left.
	\Phi\left( \frac{(i-1)\Delta+\frac{\Delta}{2}-\Delta\lfloor \frac{\theta}{\Delta} \rfloor-\Delta\langle \frac{\theta}{\Delta} \rangle}{\sigma} \right) 
	\right]^{N_i(\theta)}
\label{likelihoodA}	
\end{split}
\end{align}
where $N_i(\theta)$ is the number of occurences of the $i$--th quantization code--bin when
$\theta$ is the DC input value.
By defining $\overline{\theta} \coloneqq \nicefrac{\theta}{\Delta}$ and $j\coloneqq i-\lfloor \overline{\theta} \rfloor$ and by dividing numerator and denominator in the arguments of the cumulative distribution functions, by $\Delta$, we have 
\begin{align}
\begin{split}
l(\theta)  = 
&
	\prod_{j=-\infty}^{\infty} 
	\left[ 
	\Phi
	\left( 
		\frac{j+\frac{1}{2}	- \langle \overline{\theta} \rangle}{\overline{\sigma}} 
	\right)- 
	\right.
	\\
	& \qquad \qquad \qquad
	\left.
	\Phi\left( \frac{j-\frac{1}{2} -   \langle \overline{\theta} \rangle }{\overline{\sigma}} \right) 
	\right]^{N_{j+\lfloor \overline{\theta} \rfloor}(\theta)}.
\label{likelihapp}	
\end{split}
\end{align}
Assume $\overline{\theta}_{MLE}$ as the value of $\overline{\theta}$ that maximizes (\ref{likelihoodA}). 
To show that the MSE
%$\overline{\theta}_{MLE}$ 
only depends on $\langle \overline{\theta}\rangle$, consider the maximization of (\ref{likelihapp}) when $m\Delta$ is added to $\theta$, with $m$ integer. We have:
\begin{align}
\begin{split}
l(\theta+m\Delta)= 
& 
\prod_{j=-\infty}^{\infty} 
	\left[ 
		\Phi
		\left( 
			\frac{j+\frac{1}{2}	- \langle \overline{\theta} \rangle}{\overline{\sigma}} 
		\right)- 
		\right. 
		\\
		& \qquad \qquad 
		\left.
	\Phi\left( \frac{j-\frac{1}{2} -   \langle \overline{\theta} \rangle }{\overline{\sigma}} \right) 
	\right]^{N_{j+\lfloor \overline{\theta}+m \rfloor}(\theta+m\Delta)}.
	\label{likelihoodC}
\end{split}
\end{align}
Since with $m$ integer, $\lfloor \overline{\theta}+m \rfloor=\lfloor \overline{\theta}\rfloor+m$,
for given values of the noise sequence $\eta[0], \ldots, \eta[N-1]$, $N_{j+\lfloor \overline{\theta}+m \rfloor}(\theta+m\Delta)=N_{j+\lfloor \overline{\theta}\rfloor+m}(\theta+m\Delta)=
N_{j+\lfloor \overline{\theta}\rfloor}(\theta)$. Thus (\ref{likelihoodC}) is maximized by $\overline{\theta}_{MLE}+m$ and the quadratic error 
$(\overline{\theta}_{MLE}-\overline{\theta})^2$ does not depend on $m$ but only on $\langle \overline{\theta} \rangle$. Since this is true for any noise sequence $\eta[0], \ldots, \eta[N-1]$, also the MSE associated to MLE is periodic with $\Delta$.

\end{document}